\documentclass[printer,traditabstract]{aa}
\usepackage[varg]{txfonts}
\usepackage{graphicx,amssymb,float,hyperref,natbib}
\hypersetup{colorlinks=true,citecolor=blue,linkcolor=black,filecolor=black}
\bibpunct{(}{)}{;}{a}{}{,} % follow A&A syle

\newcommand{\msol}{$M_{\odot}$}
\newcommand{\lsol}{$L_{\odot}$}

\newcommand{\ceo}{C$^{18}$O}
\newcommand{\tco}{$^{13}$CO}

\newcommand{\mco}{$^{12}$CO}

\begin{document}

\title{Evolution of CO lines in time-dependent models of protostellar disk 
formation}

\author{D.~Harsono \inst{\ref{inst1}, \ref{inst2}} \and R.~Visser
\inst{\ref{inst3}} \and S.~Bruderer \inst{\ref{inst4}} \and
E.~F. van Dishoeck \inst{\ref{inst1}, \ref{inst4}} \and L.~E.~Kristensen
\inst{\ref{inst1},\ref{inst5}} }
\institute{
Sterrewacht Leiden, Universiteit Leiden, Niels Bohrweg 2, 2300 RA, Leiden, the
Netherlands \email{harsono@strw.leidenuniv.nl} \label{inst1}
\and
SRON Netherlands Institute for Space Research, PO Box 800, 9700 AV, Groningen,
The Netherlands \label{inst2}
\and
Department of Astronomy, University of Michigan, 500 Church Street, Ann Arbor,
MI, 48109-1042, USA \label{inst3}
\and
Max-Planck-Institut f{\" u}r extraterrestrische Physik, Giessenbachstrasse 1,
85748, Garching, Germany \label{inst4}
\and
Harvard-Smithsonian Center for Astrophysics, 60 Garden Street,
Cambridge, MA 02138, USA \label{inst5} }

\abstract{
\textit{Context}. Star and planet formation theories predict an evolution in the
density, temperature, and velocity structure as the envelope collapses and forms
an accretion disk.  While continuum emission can trace the dust evolution,
spectrally resolved molecular lines are needed to determine the physical
structure and collapse dynamics. \\
\textit{Aims}.  The aim of this work is to model the evolution of the molecular
excitation, line profiles, and related observables during low-mass star
formation.  Specifically, the signatures of disks during the deeply embedded
stage ($M_{\rm env} > M_{\star}$) are investigated.\\
\textit{Methods}. The semi-analytic 2D axisymmetric model of Visser and 
collaborators has been used to describe the evolution of the density, stellar
mass, and luminosity from the pre-stellar to the T-Tauri phase.  A full 
radiative transfer calculation is carried out to accurately determine the 
time-dependent dust temperatures. The time-dependent CO abundance is obtained 
from the adsorption and thermal desorption chemistry.  Non-LTE near-IR, FIR, 
and submm lines of CO have been simulated at a number of time steps.\\
\textit{Results}. In single dish (10--20$''$ beams), the dynamics during the
collapse are best probed through highly excited \tco\  and \ceo\ lines, which 
are significantly broadened by the infall process.  In contrast to the
dust temperature, the CO excitation temperature derived from submm/FIR data
does not vary during the protostellar evolution, consistent with \ceo\
observations obtained with {\it Herschel} and from ground-based telescopes.  The
near-IR spectra provide complementary information to the submm lines by probing
not only the cold outer envelope but also the warm inner region.  The near-IR
high-$J$ ($\ge 8$) absorption lines are particularly sensitive to the physical
structure of the inner few AU, which does show evolution.  The models indicate
that observations of \tco\ and \ceo\ low-$J$ submm lines within a
$\le$1$\arcsec$ (at 140 pc) beam are well suited to probe embedded  disks in
Stage I ($M_{\rm env}< M_{\star}$) sources, consistent with recent 
interferometric observations. High signal-to-noise ratio subarcsec resolution 
data with ALMA are needed to detect the presence of small rotationally supported 
disks during the Stage 0 phase and various diagnostics are discussed. The 
combination of spatially and spectrally resolved
lines with ALMA and at near-IR is a powerful method to probe the inner envelope
and disk formation process during the
embedded phase. }

\keywords{stars: formation - radiative transfer -  accretion, accretion disks - astrochemistry - methods : numerical }

\titlerunning{Evolution of CO lines}
\authorrunning{Harsono et al.}

\maketitle

%%%%%%%%%%%%%%%%%%%%%%%%%%%%%%%%%%%%%%%%%%%%%%%%%%%%%%%%%%%%%%%%%%%%%%%%%%%%%%%
%%%%%%%%%%%%%%%%%%%%%%%%%%%%%%%%%%%%%%%%%%%%%%%%%%%%%%%%%%%%%%%%%%%%%%%%%%%%%%%

\section{Introduction}\label{sec:intro}

The semi-analytical model of the collapse of protostellar envelopes
\citep{shu77,cm81,tsc84} has been used extensively to study the evolution of gas
and dust from core to disk and star \citep{young05,dunham10,visser09}.  Others
have explored the effects of envelope and disk parameters representative of
specific evolutionary stages on the spectral energy distribution (SED) and other
diagnostics \citep{whitney03b,robitaille06,robitaille07,crapsi08,tobin11}. 
These studies have focused primarily on the dust emission and its relation to
the physical structure.  On the other hand, spectroscopic observations toward
young stellar objects (YSOs) performed by many ground-based (sub)millimeter and
infrared telescopes also contain information on the gas structure
\citep{evans99}.  The molecular lines are important in revealing the kinematical
information of the collapsing envelope as well as the physical parameters of the
gas based on the molecular excitation.  The {\it Herschel} Space Observatory 
and the Atacama Large Millimeter/submillimeter Array (ALMA) provide new probes 
of the excitation and kinematics of the gas on smaller scales and up to higher
temperatures than previously possible. It is therefore timely to simulate the
predicted molecular excitation and line profiles within the standard picture of
a collapsing envelope.  The aim is to identify diagnostic signatures of the
different physical components and stages and to provide a reference for studies
of more complex collapse dynamics.

A problem that is very closely connected to the collapse of protostellar
envelopes is the formation of accretion disks.  The presence of embedded disks
was inferred from the excess of continuum emission at the smallest spatial
scales through interferometric observations 
\citep[eg.][]{km90,brown00,jorgensen05,prosac09}.  Their physical structure can 
be determined by the combined modeling of the SED and the interferometric 
observations \citep{jorgensen05,brinch07, enoch09}.  However, the excess 
continuum emission at small scales can also be due to other effects of a 
(magnetized) collapsing rotating envelope \citep[pseudo-disk,][]{chiang08}. 
Therefore, resolved molecular line observations from interferometers such as 
ALMA are needed to clearly detect the presence or absence of a stable rotating 
embedded disk \citep{brinch07, brinch08,lommen08,prosac09,tobin12}.

A number of previous studies have modeled the line profiles based on the
spherically symmetry inside-out collapse scenario described by \citet{shu77}
\citep[eg.,][]{zhou93,hogerheijde00,hogerheijde01,lee04,lee05,
evans05}.  Also, numerical hydrodynamical collapse models have been coupled with
chemistry and line radiative transfer to study the molecular line evolution in
1D \citep{aikawa08} and 2D \citep{brinch07,weeren09}.  Best-fit collapse
parameters (e.g., sound speed and age) are obtained but depend on the
temperature structure and abundance profile of the model. \citet{visser09} and
\citet{visser10} developed 2D semi-analytical models that describe the density
and velocity structure as matter moves onto and through the forming disk.  This
model has been coupled with chemistry \citep{visser09,visser11}, but no line
profiles have yet been simulated. The current paper presents the first study of
the CO molecular line evolution within 2D disk formation models.  CO and its
isotopologs are chosen because it is a chemically stable molecule and readily
observed.
 
Observationally, most early studies of low-mass embedded YSOs focused on the
low$-J$ ($J_{\rm u } \le 6$) (sub-)millimeter CO lines in 20--30$\arcsec$ beams,
thus probing scales of a few thousand AU in the nearest star-forming regions
\citep[e.g.,][]{ belloche02,jorgensen02,lee04,young04, crapsi05}.  
More recently, ground-based high-frequency observations of large samples up to
$J_{\rm u}$=7 are becoming routinely available
\citep{hogerheijde98,vankempen09b,vankempen09c,vankempen09a} and {\it
Herschel}-HIFI \citep{hifi} has opened up spectrally resolved observations of CO
and its isotopologs up to $J_{\rm u}=16$ ($E_{\rm u}=660$~K)
\citep{yildiz10,yildiz12}.   In addition, the PACS \citep{pacs} and SPIRE
\citep{insspire} instruments provide spectrally unresolved CO data from $J_{\rm
u}=4$ to $J_{\rm u}=50$ ($E_{\rm u}=55-7300$ K), revealing multiple
temperature components 
\citep[e.g.,][]{vankempen10a,herczeg12,goicoechea12,manoj13}.   Although the
interpretation of the higher lines requires additional physical processes than
those considered here \citep{visser12}, our models provide a reference frame
within which to analyze the lower-$J$ lines.

\begin{figure}
 \centering
  \raisebox{0.20cm}{\begin{minipage}[c]{0.98\linewidth}
\resizebox{\hsize}{!}{\includegraphics[bb=0 0 295 220,
angle=0,clip]{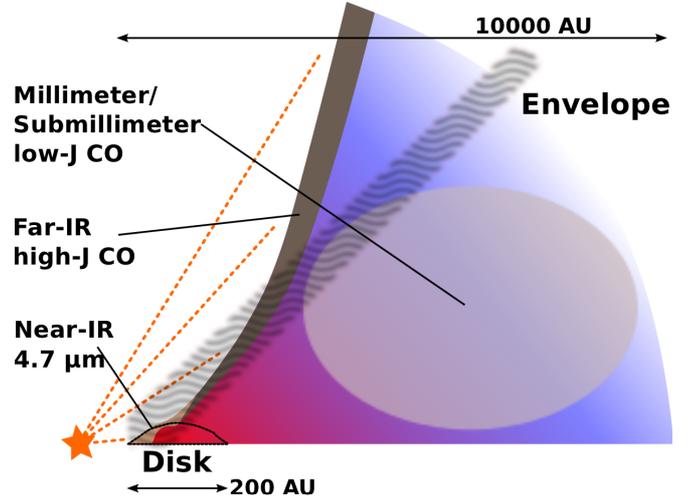}}
\end{minipage}}
\caption{Sketch of one quadrant of an embedded protostellar system with a disk 
and envelope. The FIR/millimeter/submillimeter emission comes from the 
envelope and outflow cavity walls.  On the other hand, the NIR absorption 
(gray shaded region) also probes the warm region close to the star through a 
pencil beam, illustrating the complementarity of the techniques. The dotted 
orange lines indicate the stellar light.  }
 \label{fig:cart1}
\end{figure}

Complementary information on the CO excitation is obtained from
near-IR (NIR) observations of the 4.7 $\mu$m fundamental CO (v=1--0) band seen 
in absorption toward YSOs. Because the absorption probes a pencil beam
line of sight toward the central source, the lines are more sensitive to the
inner dense part of the envelope than the submillimeter emission line data,
which are dominated by the outer envelope (see Fig.~\ref{fig:cart1}).  The
NIR data generally reveal a cold ($< 30$ K) and a warm ($> 90$ K) component
\citep{mitchell90,boogert02,brittain05,smith09,herczeg11}.  The cold component
is seen toward all YSOs in all of the isotopolog absorptions.  However,
the warm temperature varies from source to source.   The question is whether
the standard picture of collapse and disk formation can also reproduce these
multiple temperature components.

A description of the physical models and methods is given in
Section~\ref{sec:method}.  The evolution of the molecular excitation and
the resulting far-infrared (FIR) to submm lines of the pure rotational 
transitions
of CO is discussed in Section~\ref{sec:submm}, whereas the NIR
ro-vibrational transitions is presented in Section~\ref{sec:nir}.  The
implication of the results and whether embedded disks can be observed
during Stage 0 is discussed in Section~\ref{sec:dis}.  The results and
conclusions are summarized in Section~\ref{sec:sum}.

%%%%%%%%%%%%%%%%%%%%%%%%%%%%%%%%%%%%%%%%%%%%%%%%%%%%%%%%%%%%%%%%%%%%%%%%%%%%%%
%%%%%%%%%%%%%%%%%%%%%%%%%%%%%%%%%%%%%%%%%%%%%%%%%%%%%%%%%%%%%%%%%%%%%%%%%%%%%%

\section{Method}\label{sec:method}

\subsection{Physical structure}\label{sec:phstruc}

\begin{table}
\centering
 \caption{Parameters used for the three different evolutionary models.  } \label{tbl:params}
 \begin{tabular}{ l  r r r r r r } 
   \hline  
   \hline
   Model & $\Omega_{0}$ & $c_{\mathrm{s}}$ & $r_{\mathrm{env}}$ & $t_{\mathrm{acc}}$  & $M_{\mathrm{d}}$ & 
$R_{\mathrm{d}, \mathrm{out}}$\\
      & [Hz] & [km s$^{-1}$] & [AU] & [$10^{5}$ yr] & [\msol]  & [AU] \\
   \hline
   1 & $10^{-14}$ & 0.26 & 6700 & 2.5 &  0.1  & 50\\
   2 & $10^{-14}$ & 0.19 & 12000 & 6.3 &  0.2 & 180 \\
   3 & $10^{-13}$ & 0.26 & 6700 & 2.5 & 0.4  & 325 \\
   \hline
   \end{tabular}

\end{table}

\defcitealias{tsc84}{TSC84}
\defcitealias{shu77}{S77}
\defcitealias{cm81}{CM81}

The two-dimensional axisymmetric model of \citet{visser09},
\citet{visser10} and \citet{visser11} was used to simulate the collapse of a
rotating isothermal spherical envelope into a pre-main sequence star with a
circumstellar disk.  The model is based on the analytical collapse solutions of
\citet[][hereafter S77]{shu77}, \citet[][hereafter CM81]{cm81} and
\citet[][hereafter TSC84]{tsc84}.  The formation and evolution of the
disk follow according to the $\alpha_S$ viscosity prescription, which includes
conservation of angular momentum \citep{ss73,lp74}.  The dust temperature
structure ($T_{\mathrm{dust}}$) is a key quantity for the chemical
evolution, so it is calculated through full 3D continuum radiative transfer
with RADMC3D\footnote{
www.ita.uni-heidelberg.de/\textasciitilde dullemond/software/\\ radmc-3d}, considering the protostellar luminosity as the only heating source.  The gas temperature is set
equal to the dust temperature, which has been found to be a good
assumption for submm molecular lines \citep{doty02,doty04}.

The model is modified slightly in order to be consistent with observational
constraints.  The density at $r=1000$ AU ($n_{1000}$) should be at most $10^6$
cm$^{-3}$ for envelopes around low-mass YSOs \citep{jorgensen02,kristensen12},
but the interpolation scheme of \citet{visser09} violated that criterion.  
To correct this,  the \citetalias{cm81} and \citetalias{tsc84} solutions are
connected in terms of the dimension-less variable $\tau=\Omega_0 t$ by
interpolating between $100\tau^2$ and  $10\tau^2$ for $\Omega_0=10^{-13}$ Hz and
between 100$\tau^2$ and $\tau^2$ for $\Omega_0=10^{-14}$ Hz, where $\Omega_0$ is the envelope's initial solid body rotation rate.  The overall collapse, the
structure of the disk and the chemical evolution are unaffected by this
modification.  The models evolve until one accretion time, $t_{\rm acc} = M_0 /
\dot{M}$ where $\dot{M} \propto c_{\rm s}^3/G$ with $c_{\rm s}$ the initial
effective sound speed of the envelope, which is tabulated in
Table~\ref{tbl:params}.  

Three different sets of initial conditions are used in this work
(Table~\ref{tbl:params}), which are a subset of the conditions explored by
\citet{visser09} and \citet{visser10}.  Each model begins with a 1 \msol\
envelope.  The two variables $\Omega_0$ and $c_{\mathrm{s}}$ affect the final
disk structure and mass.  Model 1 ($c_{\mathrm{s}} = 0.26$ km s$^{-1}$,
$\Omega_0 = 10^{-14}$ Hz) produces a disk with a final mass ($t/t_{\rm acc}= 1$)
of 0.1 \msol\ and a final radius of 50 AU. Changing $\Omega_0$ to $10^{-13}$ Hz
(Model 3) yields the most massive and largest of the three disks (0.4 \msol, 325
AU). Starting with a lower sound speed of 0.19 km s$^{-1}$ (Model 2) produces a
disk of intermediate mass and size (0.2 \msol, 180 AU). The initial conditions
also affect the flattened inner envelope structure.  In Model 3, the flattening
of the inner envelope extends to $>300$ AU by $t/t_{\mathrm{acc}} = 0.5$ and
extends up to $\sim$1000 AU by the end of the accretion phase.  In the other two
models, the extent of the flattening is similar to the extent of the disk, with
Model 2 showing more flattening in the inner envelope than Model 1. The
different evolutionary stages are characterized by the relative
masses in the different components \citep{robitaille06}: $M_{\rm env} >>
M_{\star}$ (Stage 0, $t/t_{\rm acc} \le 0.5$), $M_{\rm env} < M_{\star}$ but
$M_{\rm env} > M_{\rm disk}$ (Stage 1, $t/t_{\rm acc} > 0.5$) and $M_{\rm env} < M_{\rm disk}$ (Stage 2) (see Appendix B).

\begin{figure}
 \centering
 \raisebox{0.20cm}{\begin{minipage}[c]{1.0\linewidth}
\resizebox{\hsize}{!}{\includegraphics[angle=0,bb=25 217 607
575, clip]{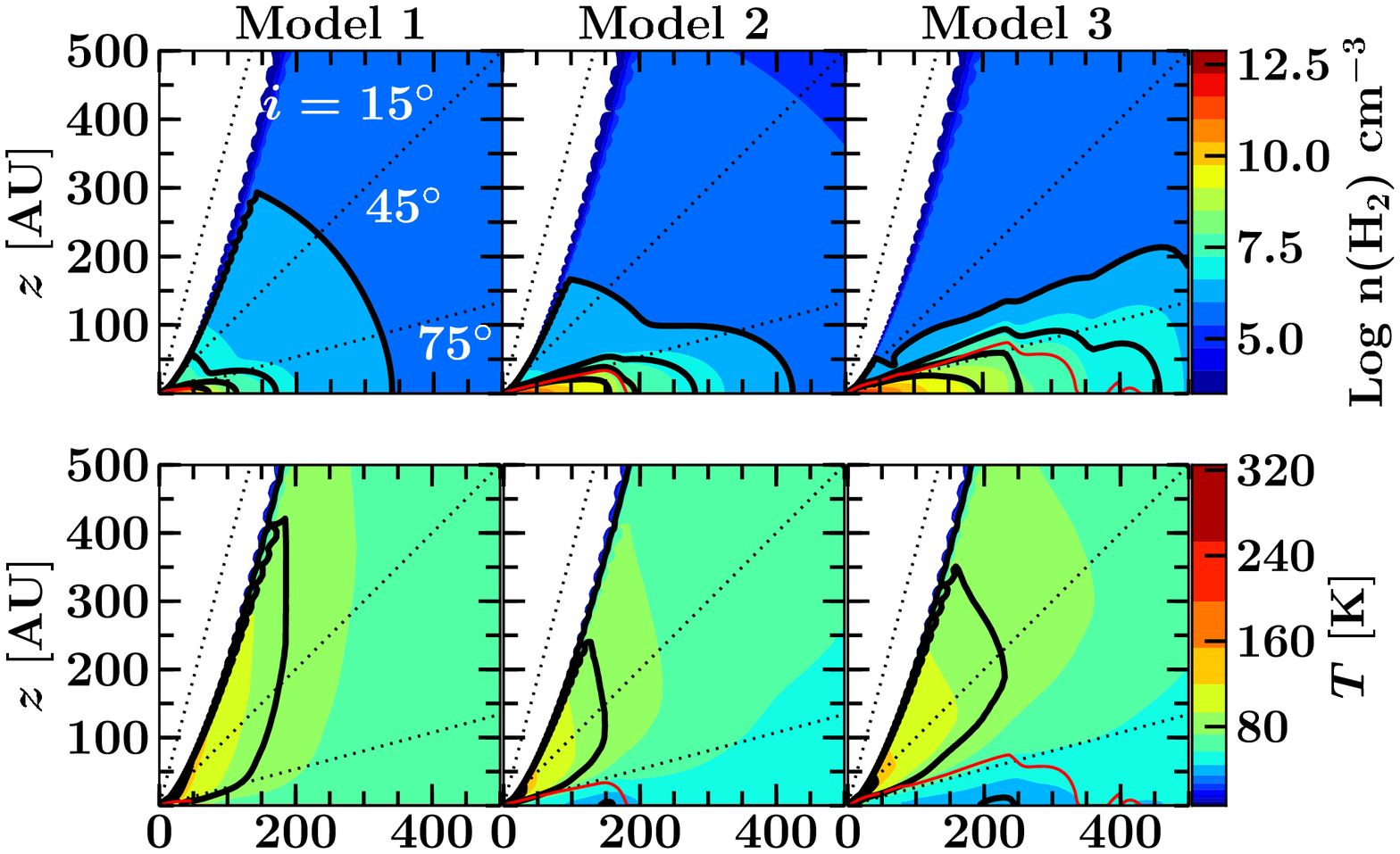}}
\end{minipage}} 
 \raisebox{0.20cm}{\begin{minipage}[c]{1.0\linewidth}
\resizebox{\hsize}{!}{\includegraphics[angle=0,bb=25 280 600
476, clip]{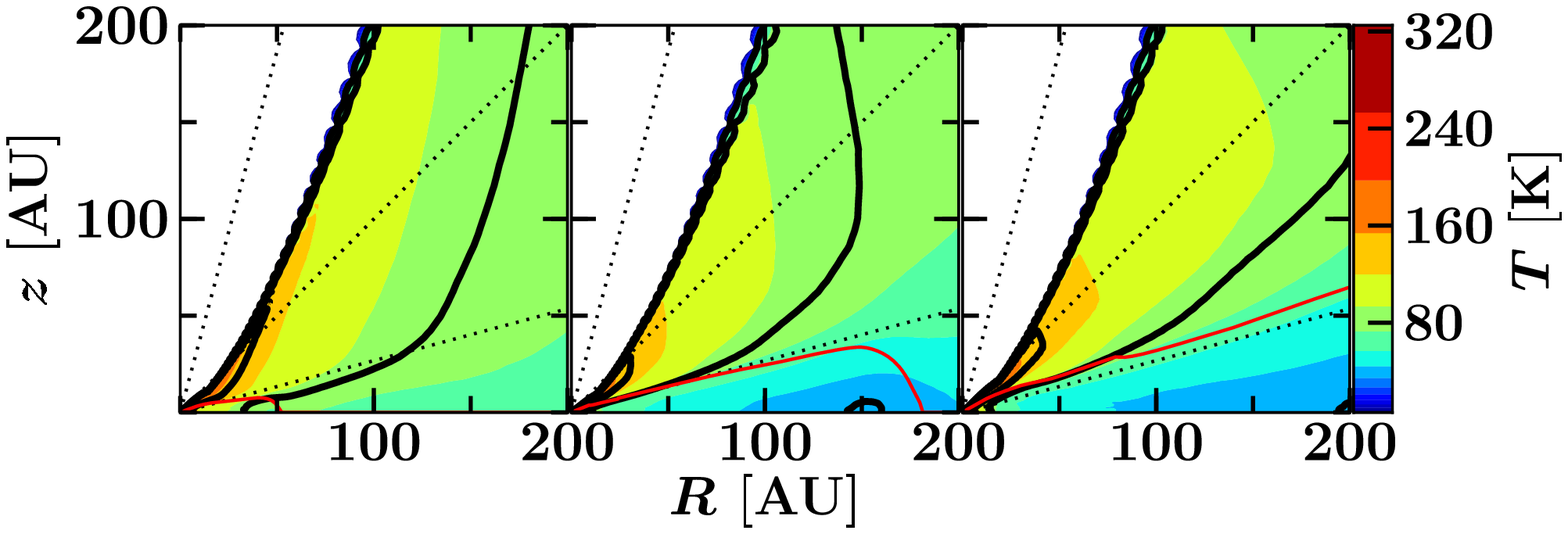}}
\end{minipage}} 
\caption{Final disk structure for the three models from Table~\ref{tbl:params}
at $t=t_{\mathrm{acc}}$.  Top: gas density in the inner 500 AU. The solid line contours mark densities of $10^{6,7,8,9}$ cm$^{-3}$.  Middle: temperature structure in the inner 500 AU.  The temperature contours are logarithmically spaced from 10 to 320 K.  The 20,
50 and 100 K isotherms are marked by the solid line contours.  Bottom: temperature structure in the inner 200 AU. In all panels, the dotted lines illustrate the lines of sight at inclinations of 15$^{\circ}$, 45$^{\circ}$ and 75$^{\circ}$ and the red line indicates the final disk surface.  }
 \label{fig:diskstruct}
\end{figure}

Our main purpose is to investigate how the molecular lines evolve
during different disk formation scenarios (Stage 0/I). Hence, the parameters
were chosen to explore the formation of three different disk structures.   
Figure~\ref{fig:diskstruct} zooms in on the final disk structures for the
different models at the accretion time ($t=t_{\mathrm{acc}}$) where the red
lines outline the disk surfaces. The angular velocities within these lines are
assumed to be Keplerian while regions outside these lines are described by the
analytical collapsing rotating envelope solutions \citep{visser09}.   The
density structure is 2D axisymmetric with a 3D velocity field within the
envelope and the disk.

\subsection{CO abundance}\label{sec:abun}

The $^{12}$CO abundance is obtained through the adsorption and thermal
desorption chemistry as described in Section~ 2.7 of \citet{visser09}.  The
main difference is that we have used both the forward and backward methods
\citep{visser11} to sample the trajectories through the disk and envelope. The
chemistry is still solved in the forward direction.  

CO completely freezes out at $T_{\mathrm{d}} \le 18$ K \citep{visser09}.  
For the majority of the time steps used for the molecular line simulations, the
dust temperature is well above 18 K everywhere except for the outer envelope
beyond $\sim$4000 AU.  However, due to the low densities within this region,
CO is still predominantly in the gas phase.  Only at early time steps, early
Stage 0, a large fraction of CO is frozen out within the inner envelope.
 Constant isotope ratios of $^{12}{\rm C}/^{13}{\rm C} = 70$ and $^{16}{\rm
O}/^{18}{\rm O} = 540$ \citep{wilson94} are used throughout the model to compute
the abundances of \tco\ and \ceo.

\subsection{Line radiative transfer}\label{sec:rt}

A fast and accurate multi-dimensional molecular excitation and radiative
transfer code is needed to obtain observables at a number of time steps.  We use
the escape probability method from \citet{bruderer12} (based
on \citealt{takahashi83} and \citealt{bruderer10}).  

The most important aspect of simulating the rotational lines is the gridding of
the physical structure. There are three components in the model that require
proper gridding (Fig.~\ref{fig:gridexample}): the outflow cavity, the envelope
($>6000$ AU) and the disk ($<300$ AU).  To resolve the steep gradients between
the different regions, 15\,000 -- 25\,000 cells are used. A detailed
description of the grid can be found in Appendix~\ref{app:A}.  The line images
are rendered at a number of time steps with a resolution of 0.1 km s$^{-1}$ to
resolve the dynamics of the collapsing envelope and the rotating disk.

In addition to pure rotational lines in the submm and FIR, this work also
explores the evolution of the NIR fundamental v=1--0 rovibrational
absorption of CO at 4.7 $\mu$m.  RADLite is used to render the
large number of NIR ro-vibrational spectra \citep{radlite}.  This is done by
assuming that all molecules are in the vibrational ground state given by the
non-LTE calculation described above.  { The assumption is valid considering 
that the observed NIR emission lines by \citet{herczeg11} are blue-shifted by a 
few km s$^{-1}$ and have broader line widths than that expected from $T_{\rm 
gas} = T_{\rm dust}$ that is being presented here.  The observed emission toward 
the YSOs seem to originate from additional physics that are not present in our 
models.  Thus, for the models presented here, it is valid to consider that the 
emission component from the inner region is negligible. }

The main collisional partners are p-H$_2$ and o-H$_2$ with the
collisional rate coefficients obtained from the LAMDA database
\citep{lamda,yang10}.  The dust opacities used in our model are a distribution
of silicates and graphite grains covered by ice mantles \citep{crapsi08}.
Finally, the gas-to-dust mass ratio is set to 100.

\subsection{Comparing to observations}\label{sec:compobs}

The molecular lines are simulated considering a source at a distance
of 140 pc.  High spatial resolution in raytracing is needed for the $J_{\rm u}
\ge 5$ lines because of the small warm emitting region.  A constant turbulent
width of $b$ = 0.8 km s$^{-1}$ is used in addition to the temperature and infall
broadening to be consistent with the observed C$^{18}$O turbulent widths toward
quiescent gas surrounding low-mass YSOs with beams $> 9\arcsec$
\citep{jorgensen02}.

The simulated spectral cubes are convolved with Gaussian beams of 1${\arcsec}$,
9${\arcsec}$ and 20${\arcsec}$ with the {\it convol} routine in the {\it
MIRIAD} data reduction package \citep{miriad}. The telescopes that observe the
low-$J$ rotational lines ($J_{\mathrm{u}} \le 5$) typically have $\geq
15{\arcsec}$  beams, as does {\it Herschel}-HIFI for the higher-$J$ lines with
$J_{\rm u}\approx 10$.  A 9${\arcsec}$ beam is appropriate for observations of
$J_\mathrm{u} > 5$ performed with, e.g., the ground-based APEX telescope at 650
GHz and with {\it Herschel}-PACS and HIFI for $J_{\rm u}>14$. A 1$\arcsec$ beam
simulates interferometric observations to be performed with ALMA.  The submm
lines are rendered nearly face-on ($i \sim 5^{\circ}$) since this is the
simplest geometry to quantify the disk contribution.  For studies of the
evolution of the velocity field, an inclination of 45$^\circ$ is taken. The
NIR lines are analyzed for 45$^{\circ}$ and 75$^\circ$ inclination to study
the different excitation conditions between lines of sight through the inner
envelope and the disk (Fig.~\ref{fig:diskstruct}).  A line of sight of
45$^{\circ}$ probes the inner envelope and does not go through the disk while
a line of sight of 75$^{\circ}$ grazes the top layers of the disk.  Both a
pencil beam approximation toward the center is taken as well as the full RADLite
simulation to study the radiative transfer effects on those lines. 

\subsection{Caveats}

The synthetic CO spectra were simulated without the presence of fore-
and background material, such as the diffuse gas of the large-scale
cloud where these YSOs are forming.  The overall emission of the
large-scale cloud can affect the observed line profiles within a large
beam ($> 15\arcsec$), in particular the $J_{\mathrm{u}} \le 4$ lines
of \mco\ and \tco\ and the $J_{\mathrm{u}} \le 2$ lines of \ceo.  Also
not included in the simulations are energetic components such as jets,
shocks, UV heating and winds.  These energetics strongly affect the \mco\
lines, especially the intensity of the $J_{\rm u} \ge 6$ rotational transitions
\citep{spaans95, vankempen09b,visser12}.  Luminosity flares associated with
episodic accretion events can affect the CO abundance structure as discussed in
\citet{visser12b}.  Their effect on the evolution of the CO line profile and excitation is beyond the scope of this paper.  In general, a higher
CO flux is expected during an accretion burst. 

%%%%%%%%%%%%%%%%%%%%%%%%%%%%%%%%%%%%%%%%%%%%%%%%%%%%%%%%%%%%%%%%%%%%%%%%%%%%%%%
%%%%%%%%%%%%%%%%%%%%%%%%%%%%%%%%%%%%%%%%%%%%%%%%%%%%%%%%%%%%%%%%%%%%%%%%%%%%%%%

\section{FIR and submm CO evolution}\label{sec:submm}

The FIR and submm CO lines up to the $10-9$ transition
($E_{\mathrm{u}}$ = 290--304 K) have been simulated with $i$=5$^\circ$.
The geometrical effects on the line intensities (\tco\ and \ceo) are less than
30\% for different inclinations and the derived excitation temperatures differ
by less than 5\%, which is smaller than the $rms$ error on the derived
temperatures. On the other hand, inclination strongly affects the derived
moment maps and interpretation of the velocity field, therefore an inclination
of 45$^{\circ}$ is used in Section~\ref{sec:evolvels}.

\subsection{Line profiles}\label{sec:lines}

\begin{figure}
 \centering
  \raisebox{0.25cm}{\begin{minipage}[c]{1.00\linewidth}
\resizebox{\hsize}{!}{\includegraphics[angle=0,bb=35 220 540 595,
clip]{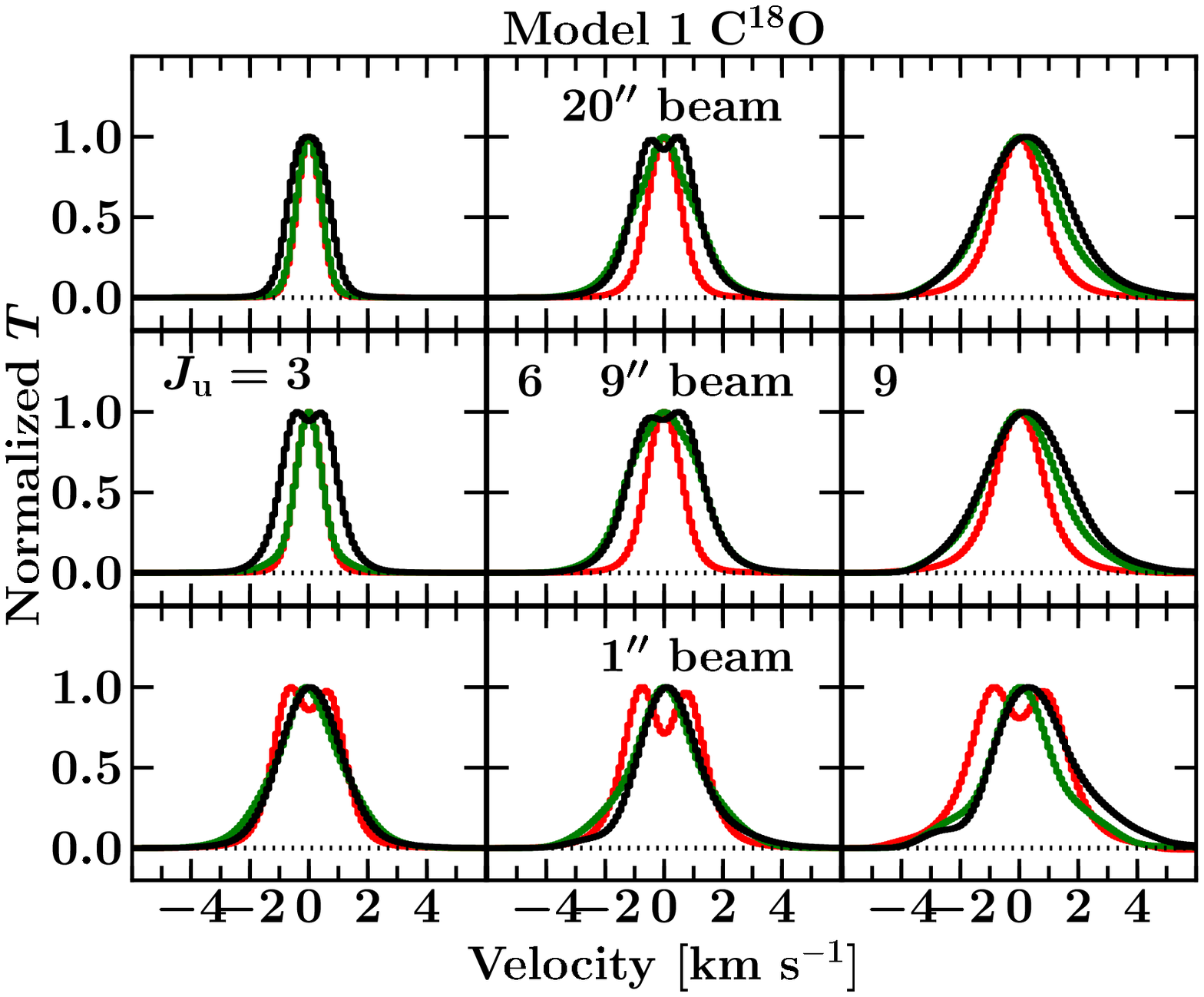}}
\end{minipage}} 
\caption{ Normalized \ceo\ line profiles as functions of evolution
for $J_{\mathrm{u}} = 3$, 6 and 9 at $t/t_{\mathrm{acc}} = 0.13$ (red), 0.50
(green) and 0.96 (black) for $i = 5^{\circ}$ orientation for Model 1.  The lines
are convolved to beams of 20$\arcsec$ (top), 9$\arcsec$ (middle) and 1$\arcsec$
(bottom).}
 \label{fig:c18olines}
\end{figure}

The first time step used for the molecular line simulation is at $t = 1000$
years ($t/t_{\rm acc} \sim 10^{-3}$), where the central heating is not yet
turned on.  The evolution of the mass and the effective temperature is shown in
the Appendix~\ref{app:B}.  The effective temperature during this time step is 10
K and the models are still spherically symmetric.  The \citetalias{tsc84}
velocity profile for this time step reaches a maximum radial component of 10 km
s$^{-1}$ at 0.1 AU.  However, CO is frozen out in the inner envelope, so the FIR
and submm lines (populated up to $J_{\rm u} \le 8$) show neither wing emission
nor blue asymmetry as typical signposts of an infalling envelope. Instead, the
CO lines probe the static outer envelope at this point, where the low density
has prevented CO from freezing out.

\begin{figure}
 \centering
  \raisebox{0.25cm}{\begin{minipage}[c]{1.00\linewidth}
\resizebox{\hsize}{!}{\includegraphics[angle=0,bb=45 187 562
478,clip]{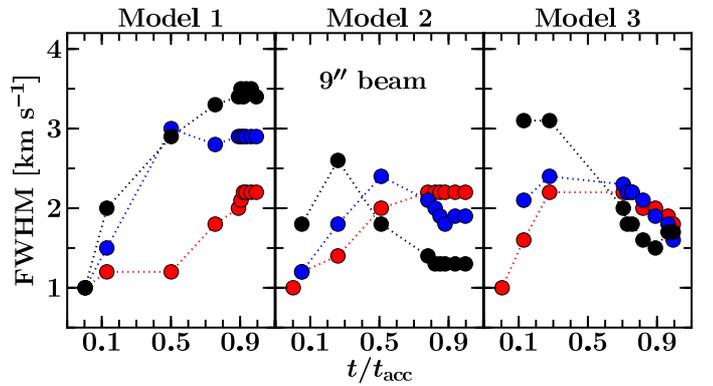}}
\end{minipage}} 
\caption{ \ceo\ 3-2 (red), 6-5 (blue) and 9-8 (black) FWHM evolution within a
9$\arcsec$ beam for the 3 different models
at $i = 5^{\circ}$
orientation.}
 \label{fig:allc18ofwhm}
\end{figure}

The line profiles change shape once the effective source temperature
increases to above a few thousand K at $t > 10^{4}$ yr (see
Fig.~\ref{fig:c18olines} and Appendix \ref{app:B} for the line profile
evolution).  The line profiles become more asymmetric (Fig.~\ref{fig:allc18o9})
toward higher excitation and smaller beams, because in both cases a larger
contribution from warm, infalling gas in the
inner envelope and optical depth affect the emission lines.  The line widths are
due to a combination of thermal broadening, turbulent width and velocity
structure, but examination of models without a systematic velocity field confirm
that the increase in broadening is mostly due to infall \citep[see
also][]{lee04}.  Models 2 and 3 show narrower lines during Stage I ($t/t_{\rm
acc} \ge 0.5$) since they are relatively more rotationally dominated than in
Model 1.  As shown in Fig.~\ref{fig:allc18ofwhm}, the higher-$J$ ($J_{\rm u} \ge
5$) lines are broadened significantly during Stage 0.

The line profiles in a 1$\arcsec$ beam depend strongly on the velocity
field on small scales and show more structure than the line profiles
in the larger beams (Figs.~\ref{fig:c18olines} and \ref{fig:evollines}). The
lines in a 1$\arcsec$ beam are broader and show multiple velocity components (in
particular for \mco\ and \tco)  reflecting the complex dynamics due to infall
and rotation plus the effects of optical depth and high angular resolution.  A
combination of CO isotopolog lines can provide a powerful diagnostic of the
velocity and density structure on $10-1000$ AU scales (see also \S 3.4).

In summary, the infalling gas can significantly broaden the optically thin
gas as shown in Fig.~\ref{fig:allc18ofwhm}.  The velocity profile in the inner
envelope affects the broadening of the high-$J$ ($J_{\rm u} > 6$) line.  These
lines can be a powerful diagnostic of the velocity and density structure in the
inner 1000 AU ($\le 9\arcsec$ beams).

\subsection{CO rotational temperature}\label{sec:trots}

The observed integrated intensities (K km s$^{-1}$) are commonly analyzed by
constructing a Boltzmann diagram and calculating the associated rotational
temperature.  Another useful representation is to plot the integrated flux
($\int F_{\nu} d\nu$ in W m$^{-2}$) as a function of upper level rotational
quantum number (spectral line energy distribution or SLED) to determine the
$J$ level at which the peak of the molecular emission occurs.  The conversion
between integrated intensities and integrated flux is given by 
\begin{equation}
 \int F_{\nu} d\nu = \frac{2 k}{\lambda^3} d\Omega \int T_{v} dv.
\end{equation}

\begin{figure}
\raisebox{0.20cm}{\begin{minipage}[c]{1.00\linewidth}
\resizebox{\hsize}{!}{\includegraphics[angle=0,bb=35 180 540 612,
clip]{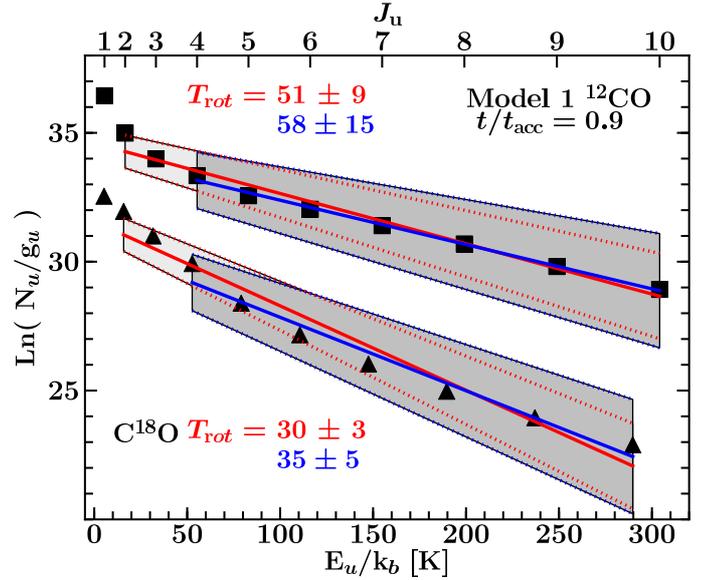}}
\end{minipage}}
\caption{Examples of single-temperature fits through the \mco\
  (squares) and \ceo\ (triangles) lines obtained from Model
  1 fluxes at $t/t_{\rm acc}=0.9$ in a 9$''$ beam.  The red line shows the
  linear fit from $J_{\rm u}=2$ up to $J_{\mathrm{u}} = 10$ while the
  blue line is the fit up starting from $J_{\rm u} =4$ up to $J_{\rm u} = 10$. 
The gray shaded regions
  indicate one standard deviation from the best fit.}
 \label{fig:example-trots}
\end{figure}

\begin{figure}
\raisebox{0.20cm}{\begin{minipage}[c]{1.00\linewidth}
\resizebox{\hsize}{!}{\includegraphics[angle=0,bb=35 180 540
590,clip]{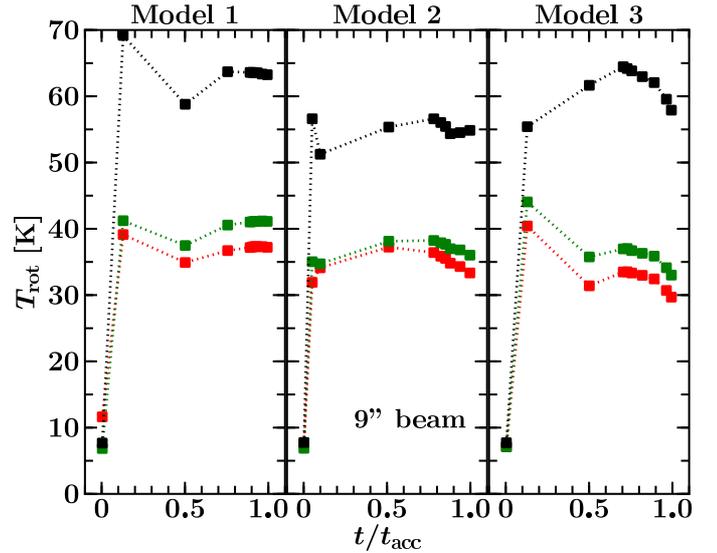}} 
\end{minipage}}
\caption{Evolution of the rotational temperature ($T_{\mathrm{rot}}$) from a
single-temperature fit through $J_{\mathrm{u}} = 2-10$.  The different colors
are for different isotopologs: $^{12}$CO (black), $^{13}$CO (green) and
C$^{18}$O (blue).  The typical errors are 15, 5 and 3 K for \mco, \tco\ and
\ceo, respectively.  The different panels show the results for different models
as indicated.}
\label{fig:evoltrot}
\end{figure}

Examples of single-temperature fits through the modeled lines are shown in
Fig.~\ref{fig:example-trots} for Model 1.  Observationally, temperatures are
generally measured from $J_{\rm u} = 2$ to 10 \citep{yildiz12,yildiz13a}
and from 4 to 10 \citep{goicoechea12}.  The evolution of the rotational
temperatures within a 9$\arcsec$ beam is shown in Fig.~\ref{fig:evoltrot} for
the three models.  At very early times, $t/t_{\rm acc} < 0.1$, a single
excitation temperature of $\sim$ 8--12 K characterizes the CO Boltzmann
diagrams. After the central source turns on, the excitation temperature is
nearly constant with time.  In a beam of 9$\arcsec$, most of the flux up to
$J_{\rm u} = 10$ comes from the $>$ 100 AU region, which is not necessarily the
$>100$ K gas \citep{yildiz10} even for face-on orientation.  Therefore, within
$\ge 9\arcsec$ beams, the observed emission is not sensitive to the warming up
of material as the system evolves.

A value of 46 to 65 K characterizes the \mco\ distribution for all models. 
However, \mco\ is optically thick, which drives the temperature to higher values
due to under-estimated low-$J$ column densities.  The \ceo\ and \tco\
distributions are fitted with similar excitation temperatures between 29 and 42
K within a 9$\arcsec$ beam.  Because Model 3 has a higher inner envelope
density, the derived rotational temperature at the second time step is
relatively high ($\gtrsim$ 40 K) due to an optical depth effect.

\begin{figure}
\centering
\raisebox{0.20cm}{\begin{minipage}[c]{1.00\linewidth}
\resizebox{\hsize}{!}{\includegraphics[angle=0,bb=35 180 540 590,
clip]{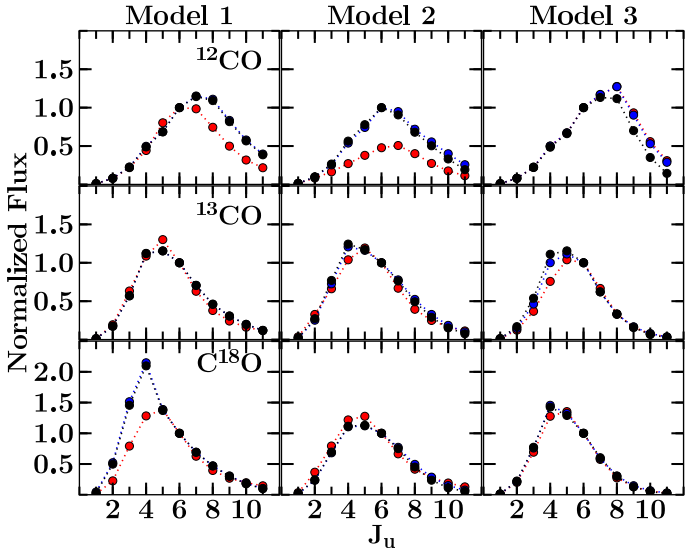}}
\end{minipage}} 
\caption{ Evolution of normalized CO spectral line energy distribution (SLED) in
a 9$\arcsec$ beam.  The CO integrated fluxes (W m$^{-2}$) are normalized to
$J_{\mathrm{u}} =6$.  The different colors indicate different times: $t/t_{\mathrm{acc}} \sim 0.50$ (red), 0.75 (blue),
0.96 (black).}
  \label{fig:coladder}
 \end{figure}

 The integrated fluxes can also be used to construct CO SLEDs.  The \mco\
SLED peaks at $J_{\mathrm{u}} = 6-8$ with small dependence on the heating
luminosity (Fig.~\ref{fig:coladder}). The peak of the SLED {\it does not}
depend on the beam size for beams $\ge 9{\arcsec}$.  The same applies to the
 \tco\ and \ceo\ SLEDs which peak at $J_{\mathrm{u}} = 4-5$.  There is no
significant observable evolution in the SLEDs once the star is turned on. 
A passively heated system is characterized by such SLEDs regardless of
 the evolutionary state as long as the heating luminosity is  $\sim 1-10$ \lsol\
($T_{\star} \sim$ 1000--4500 K).

The picture within a 1$\arcsec$ beam is different since it resolves
 the inner envelope.  The best-fit single temperature component
 depends on whether or not the disk fills a significant fraction of the beam. 
If it does (Models 2 and 3), the \ceo\ and \tco\ rotational temperatures are
$\sim$40--50 K.  There is less spread than within the 9$\arcsec$
beam due to the higher densities bringing the excitation closer to LTE. 
Without a significant disk contribution in a $1 \arcsec$ beam (Model 1), the
\tco\ excitation temperature is closer to 60 K while that of \ceo\ is similar
to the other models at $\sim 40$ K.  The \mco\ rotational temperature is 
$T_{\rm rot} \simeq 70$ K in all of the models.  Furthermore, the CO SLED within
a 1$\arcsec$ beam peaks at $J_{\rm u} = 6, 8$ and  $10$ for \ceo, \tco\ and
\mco, respectively.

 A comparison with the sample presented in \citet{yildiz13a} indicates that the 
lack of $T_{\rm rot}$ evolution is consistent with observations, and our 
predicted values for $T_{\rm rot}$ match the data.  On the other hand, the 
observed \tco\ $T_{\rm rot}$ and SLED are generally higher than predicted, 
which suggests an additional heating component is needed to excite the \tco\ 
lines.  A more detailed comparison between model prediction and observation can 
be found in \citet{yildiz13a}.

  In summary, both rotational temperatures derived from Boltzmann
 diagrams and CO SLEDs of passively heated systems do not evolve with
 time once the star is turned on.  Optically thin \tco\ and \ceo\
 lines are characterized by single excitation temperatures of the
 order 30--40 K within a wide range of beams ($\ge 9 \arcsec$).  The CO
 SLED peaks at $J_{\mathrm{u}} = 7\pm 1$ and $J_{\mathrm{u}} = 4-5$
 for \mco\ and other isotopologs, respectively.  In $1 \arcsec$ beam, the
peak of the SLED shifts upward by $\sim 2$ $J$ levels and a warmer rotational
temperature by 10 K in the presence of a disk.

\subsection{Disk contribution to line fluxes}\label{sec:diskcon}

After deriving a number of observables, an important question is, what the
fraction of the flux contributed by the disk is?  This contribution can be 
calculated from the averaged (over line profile and direction) escape 
probability of the line emission $\eta_{\rm ul}$, which is the probability that 
a photon escapes both the dust and line absorption { \citep{takahashi83, 
bruderer12}. The escape probabilities are used to calculate the cooling rate, 
$\Gamma_{\rm cool, \nu}$, of each computational cell and each transition with 
the following formula 
\begin{equation}
 \Gamma_{\rm cool, \nu} = \frac{V}{4 \pi} h \nu_{\rm ul} A_{\rm ul} n x_{\rm u} 
\eta_{\rm ul},
\end{equation}
where $\nu_{\rm ul}$ is the frequency of the line, $A_{\rm ul}$ is the Einstein $A$ coefficient, $x_{\rm u}$ is the normalized population level and $n$ is the density of the molecule.  The disk contribution is then the ratio of the sum of cooling rates from the cells in the disk compared to the cells within a Gaussian beam.   The cooling rates are weighted with a Gaussian beam of size 9$\arcsec$ or 1$\arcsec$ while the disk emitting region is defined as shown in Fig.~\ref{fig:diskstruct}.  The flux is then given through an integration of line of sight, $F_{\nu} = \int \Gamma_{\rm cool, \nu} ds_{\rm LOS}$,  which translates into the same constant factor in both disk and total fluxes.  }

\begin{figure}
 \centering
   \raisebox{0.20cm}{\begin{minipage}[c]{1.00\linewidth}
\resizebox{\hsize}{!}{\includegraphics[angle=0,bb=35 180 540
590, clip]{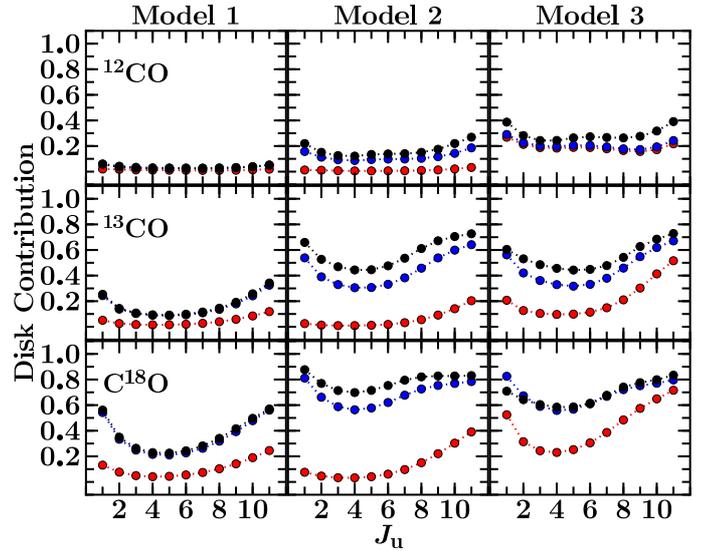}}
\end{minipage}} 
\caption{ Disk contribution as a function of $J_{\mathrm{u}}$ within
1${\arcsec}$ (140 AU at 140 pc) for the three different models.  The different
colors correspond to different evolutionary stages: $t/t_{\mathrm{acc}} \sim
0.50$ (red, $M_{\rm env}/M_{\rm disk} \ge 3$), 0.75 (blue, $M_{\rm env}/M_{\rm
disk} \sim 1$), 0.96 (black, $M_{\rm env}/M_{\rm disk} < 1$).  }
 \label{fig:almadiskcont}
\end{figure}

For a single-dish 9$\arcsec$ beam, one needs to go to higher rotational
transitions with $J_{\mathrm{u}} > 6$ at later stages to obtain a $>$ 50\% disk
contribution, if the disk can be detected at all (Fig.~\ref{fig:evoldiskcont}). 
The disk is difficult to observe directly in \mco\ and \tco\ emission since the
lines quickly become optically thick unless lines with $J_{\rm u}>10$ are
observed.  A larger disk contribution is seen in the optically thin \ceo\ lines, but here the low absolute flux may become prohibitive.

For example, the expected \ceo\ disk fluxes for $J_{\rm u} > 14$
within the PACS wavelength range are $\le 10^{-20}$ W m$^{-2}$ which will take
$> 100$ hours to detect with PACS.  {\it Herschel}-HIFI is able to spectrally
resolved the \ceo\ $J_{\rm u} = 10$ and 9 lines but has a $\sim 20$\arcsec\ beam, which lowers the disk fraction by a factor of 2 relative to a 9$\arcsec$ beam.  Peak
temperatures of only 1--8 mK during Stage 0 and 2--12 mK in Stage I phase are
expected, which are readily overwhelmed by the envelope emission.   The spectrally resolved \ceo\ spectra observed with HIFI indeed do not show any
sign of disk emission, consistent with our predictions \citep{sanjose-garcia13, yildiz12, yildiz13a}. 

A more interesting result is the disk contribution to the CO lines within a
1$''$ (140 AU) region as shown in Fig.~\ref{fig:almadiskcont}.  The simulations
suggest a significant disk contribution ($>$ 50\%) for the $^{13}$CO and
C$^{18}$O lines within this scale, even for the $J =1-0$ transition.  For Model
3, the high disk contribution is expected to happen before the system enters
Stage 1 ($M_{\mathrm{env}} \le M_{\star}$): a significant disk contribution is
seen even at $t/t_{\mathrm{acc}} = 0.3$.  However, Model 2 shows a significant
disk contribution only in the second half of the accretion phase.  Model 2 shows
a significant jump in the disk contribution between $t/t_{\rm acc} =0.5$ and
0.75 which is due to significant contribution from the flattened inner envelope
within 1$\arcsec$ at $t/t_{\rm acc} = 0.5$.  Thus, the relatively more optically
thin CO isotopolog emission is dominated by the central rotating disk starting from the Stage 1 phase ($t/t_{\rm acc} \ge 0.5$).

\begin{figure*}
 \centering
  \begin{tabular}{cc}
   \raisebox{0.20cm}{\begin{minipage}[c]{0.5\linewidth}
\resizebox{\hsize}{!}{\includegraphics[angle=0,bb=213 240 473 454,
clip]{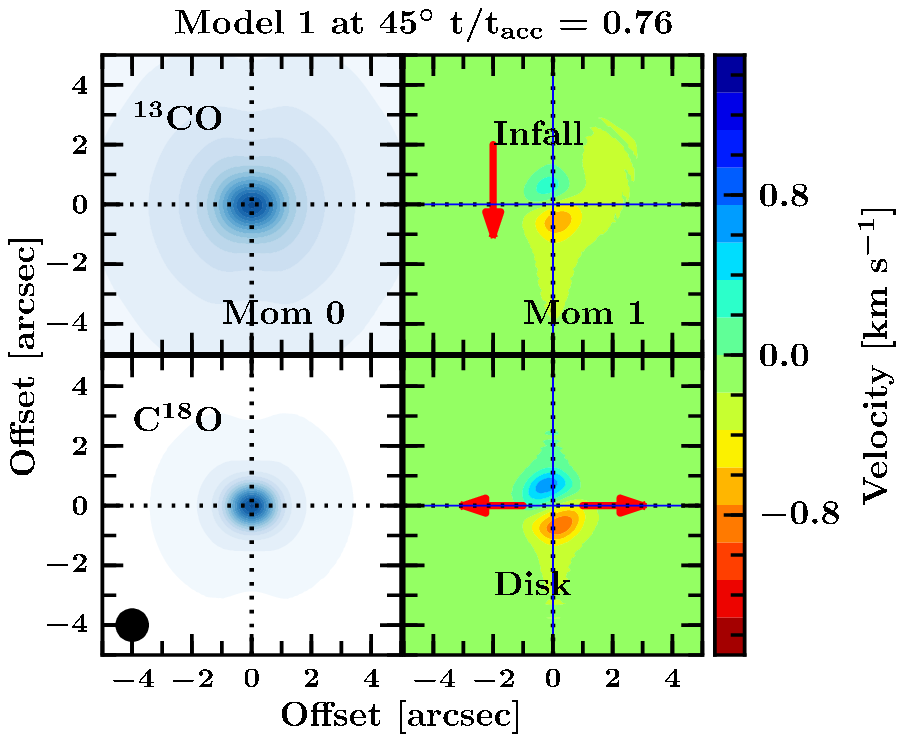}}
\end{minipage}} &
\raisebox{0.15cm}{\begin{minipage}[c]{0.35\linewidth}
\resizebox{\hsize}{!}{\includegraphics[angle=0,bb=245 242 425 454,
clip]{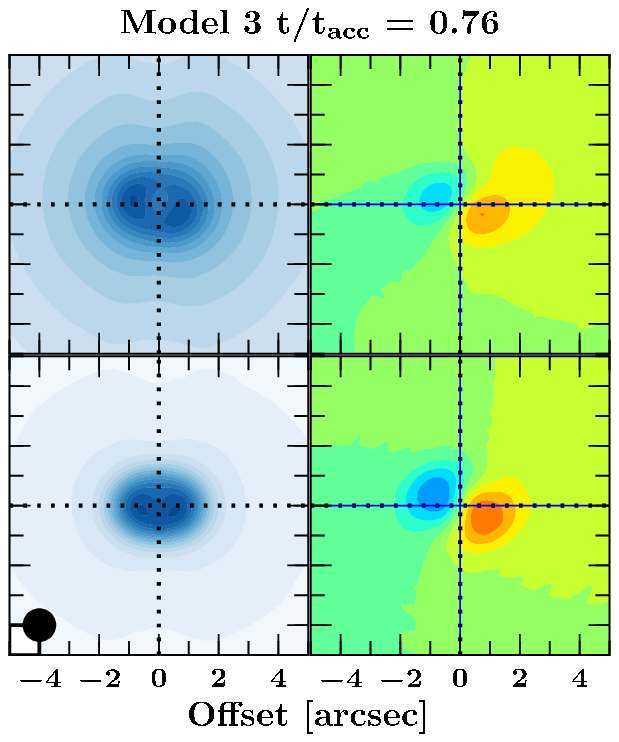}}
\end{minipage}} 
 \end{tabular}
\caption{\tco\ and C$^{18}$O zeroth (left panels) and first (right
panels) moment maps for Model 1 and 3 at 45$^{\circ}$ inclination at a
resolution of 1$\arcsec$.  The red arrows indicate the direction of infall and
rotationally supported disk.  The velocity scale is given on the right-hand
side of the left figure. }
 \label{fig:mom1map}
\end{figure*}

\subsection{Disentangling the velocity field}\label{sec:evolvels}

How can the rotating and infalling flattened envelope be disentangled
from the Keplerian motion of the disk? The evolution of the rotationally
dominated region is consistent with that reported by \citet{brinch08} based on
more detailed hydrodynamics simulations. All models become rotationally
dominated within 500 AU at $t/t_{\rm acc} > 0.3$.

\begin{figure}
 \centering
\raisebox{0.20cm}{\begin{minipage}[c]{0.95\linewidth}
\resizebox{\hsize}{!}{\includegraphics[angle=0,bb=214 236 425 456,
clip]{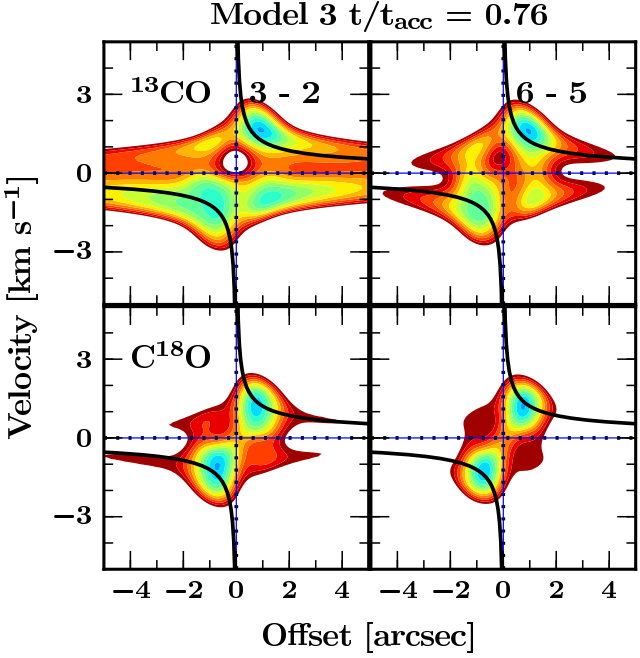}}
\end{minipage}}
\caption{Position-velocity slice for the \tco\ and \ceo\ $J_{\rm u}=3$ and
$J_{\rm u} =6$ transitions along the major axis of the zeroth moment map. The
solid black lines are the Keplerian velocity structure calculated from the
stellar mass. }
 \label{fig:pvslice}
\end{figure}

The presence of an embedded disk can be inferred from the presence of elongated
\tco\ and \ceo\ integrated intensity maps (moment 0 of Fig.~\ref{fig:mom1map})
coupled with the moment 1 map.  In the presence of a stable disk, the zeroth
moment map is elongated perpendicular to the outflow axis with a double peaked
structure.  This feature is not seen in Model 1 (Fig.~\ref{fig:mom1map} left) since the disk is much smaller than the 1$\arcsec$ beam. In the case of Model 3, the relatively massive disk exhibits a double peaked zeroth moment map which is perpendicular to the outflow direction.

A velocity gradient is seen in the moment 1 maps for both Models 1 and 3.  With
the high resolution of the modeled spectra, it is possible to differentiate
between the disk and envelope \citep{hogerheijde01}.  The presence of a velocity
gradient along the major axis of the elongation in the moment 1 map is a clear
sign of a stable embedded disk in the case of Model 3.  In addition, from the
analysis in Sect.~\ref{sec:diskcon}, we can also attribute the bulk of optically
thin emission in Model 3 to the disk.  Meanwhile, the infalling rotating
envelope contribution can be detected through the fact that the moment 1 map is
not perfectly aligned but skewed. On the other hand, a velocity gradient without
the presence of elongation in the zeroth moment map such as in Model 1
(Fig.~\ref{fig:mom1map}) indicates an infalling envelope. 

Position-velocity (PV) diagrams provide another way to study the velocity
structure. Figure~\ref{fig:pvslice} shows PV slices following the direction of
the major axis of the disk (PA = 90$^{\circ}$) in the integrated intensity map
in the 3--2 and 6--5 lines.  A simple way to analyze PV diagrams is to separate
the diagram into four quadrants and examine which quadrants have considerable
emission.  An infall dominated PV diagram without rotation is symmetric about
the velocity axis with all four quadrants filled \citep{ohashi97b}.  The
presence of rotation breaks the symmetry in the PV diagram and causes the
peak positions to be off-centered.   As found previously in hydrodynamical
simulations by \citet{brinch08}, infall dominates the PV diagrams at early times
while rotation dominates the later times.  These features have been
observed toward low-mass YSOs which suggests that generally the infalling
rotating envelope dominates the PV diagrams
\citep[eg.][]{sargent87, hayashi93,saito96, ohashi97b,hogerheijde01}.

Focusing on Model 3 at $t/t_{\rm acc} = 0.76$ (near the end of Stage I) as shown
in Fig.~\ref{fig:pvslice}, the \ceo\ PV diagrams show a clearer pure rotation
dominated structure than the \tco\ lines.  They can thus be used to constrain
the stellar mass as long as the inclination is known.  Also, the higher-$J$
lines provide a better view into the rotating structure and should give tighter
constraints on the stellar mass.  

\begin{figure}
 \centering
\raisebox{0.20cm}{\begin{minipage}[c]{0.95\linewidth}
\resizebox{\hsize}{!}{\includegraphics[angle=0,bb=135 230 482 458,
clip]{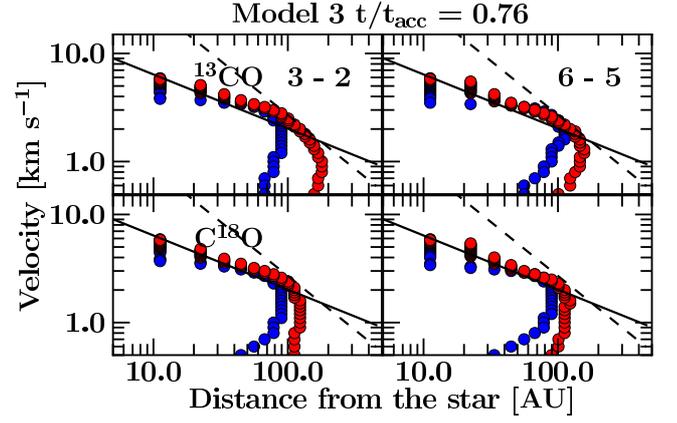}}
\end{minipage}}
\caption{Velocity as function of distance of the peak intensity position from
the protostar, as measured in each channel map and for the same transitions as
in Fig.~\ref{fig:pvslice}.  The red and blue colors represent the red- and
blue-shifted components.  The solid black line indicates the expected Keplerian
structure from the stellar mass in the model, where as the dashed lines indicate
the $v \propto R^{-1}$ relation as a comparison.}
 \label{fig:pv2}
\end{figure}

Another method is to plot the velocity as function of distance of the position
of the peak intensity \citep{sargent87}.  This is done for the red-shifted
and blue-shifted components, plotted in Fig.~\ref{fig:pv2}.  The method is
similar to the spectroastrometry technique employed in the optical and NIR
observations \citep{takami01, baines06,pontoppidan08}.  A Keplerian disk is
characterized by a $v\propto r^{-0.5}$ with large contribution from the high
velocity gas which is optically thin.  As noted by \citet{sargent87}, as
long as the line is spectrally resolved ($dV = 0.1$km s$^{-1}$), the peak
position corresponds to the maximum radius of a given velocity.  On the other
hand, the emission at low velocities ($dV \sim V$) is relatively more optically
thick and is dominated by the infalling rotating envelope which peaks closer
to the center (no offset).  Such an analysis can be performed directly from the
interferometric data and is a powerful tool in searching for embedded
rotationally stable disks out of the rotating infalling envelope.

Why is there a difference between the red-shifted and blue-shifted components 
{in Fig.~\ref{fig:pv2}}?  At high velocities, this is due to unresolved 
emission hence the different velocity components simply peak at the same 
position and the difference
reflects the uncertainty in locating the emitting region. At low velocities,
the optically thick infalling rotating envelope affects the peak positions.  In
an infalling envelope, the blue-shifted emission is relatively more optically
thin than the red-shifted emission \citep{evans99}.  Such difference in the
optical depth causes the red- and blue-shifted emissions to be asymmetric as
found in the moment 1 map.

%%%%%%%%%%%%%%%%%%%%%%%%%%%%%%%%%%%%%%%%%%%%%%%%%%%%%%%%%%%%%%%%%%%%%%%%%%%%%%%
%%%%%%%%%%%%%%%%%%%%%%%%%%%%%%%%%%%%%%%%%%%%%%%%%%%%%%%%%%%%%%%%%%%%%%%%%%%%%%%

\section{NIR CO absorption lines}\label{sec:nir}

A complementary probe of the molecular excitation conditions is
provided by the NIR CO ro-vibrational absorption lines. In this
work, we concentrate on the $\mathrm{v} = 1-0$ band at 4.76
$\mu$m. This absorption takes place along the line of sight through
the envelope and/or disk up to where the continuum is formed.  The
absorption lines are therefore computed for different inclinations of
45$^{\circ}$ and 75$^{\circ}$.  The inclinations were chosen such that
they probe the envelope ($> 15 ^{\circ}$) and the part of the disk
that is not completely optically thick such that there is enough
observable NIR continuum ($ \le 75^{\circ}$), i.e., lines of sight
that graze the disk atmosphere.  Since our focus is on the excitation,
the absorption lines have been calculated without any velocity field
besides a turbulent width (Doppler $b$) of 0.8 km s$^{-1}$.  More importantly,
the formation of the 4.7$\mu$m continuum
is strongly dependent on the inclination which affect the molecular absorption
lines.

\subsection{Evolution of NIR absorption spectra} \label{sec:nirabs}

\subsubsection{Radiative transfer and non-LTE effects}

The line center optical depth is one of the quantities derived from
the model that can be compared to observations.  This optical depth
can either be determined by computing the line of sight integrated
column densities and converting this to optical depth or by
using a full RADLite calculation.  For both approaches, the same level
populations are used, i.e., the same model for the CO excitation is
adopted as in Section 3.  In the simplest method, the line center optical
depth is obtained from
\begin{equation}
 \tau_0 = \frac{ c^3 g_\mathrm{u} }{8 \pi \sqrt{\pi} b \nu^3 g_{\mathrm{l}} }
A_{\mathrm{ul}} N_{\mathrm{l}},
 \label{eq:tau0}
\end{equation}
where $\nu$ is the line frequency and $N_{\mathrm{l}}$ is the lower level column
density along
the line of sight.  The main difference between the methods is that
RADLite solves the radiative transfer equation along the line of sight
and thus accounts for continuum and line optical depth effects and
scattered continuum photons, while such effects are not considered in
the column density approach.  In addition, RADLite accounts for the continuum
formation, while a ray through the center does not.  Thus, the two methods
effectively compare the total mass present along a ray and the observed mass as
probed by the line optical depth returned by RADLite.  The optical depth is
extracted from the RADLite spectra using the line-to-continuum ratio at the line
center.

As discussed and illustrated extensively in Appendix~\ref{app:C1}, the two
approaches can give very different results, especially for the higher
$J$ lines for which the line optical depths can differ by more than an
order of magnitude. The main reason is illustrated in
Fig.~\ref{fig:contimage}, which shows the region where the NIR
continuum arises. For small inclinations, the continuum is essentially
point-like, but for higher inclinations larger radii contribute
significantly and the continuum is no longer a point source. Thus, for
high inclinations, the high-$J$ lines start absorbing off-center, away
from warm gas in the inner few AU that are included in the column
method.  The location of the continuum is typically at $>$10 AU at $i
= 75^{\circ}$ which results in two to three orders of magnitude
difference in total column density.  For the case of $i\sim
45^{\circ}$, the result depends on the physical structure in the inner
few AU but the absorption can miss the warm high density region close
to the midplane of the disk where most of the $J \ge 10$ is located
(Fig.~\ref{fig:zoom}).

Figure~\ref{fig:nirtau0} and \ref{fig:nirtau0a} also compare LTE and non-LTE 
spectra.  Consistent with Section 3, significant differences are found for the 
higher pure rotational levels which are subthermally excited. However, the 
radiative transfer effect is generally more important than the non-LTE effects 
for viewing angles through the disk/inner envelope surface.

\begin{figure}
 \centering
% %  if referee format scale this one down
% \raisebox{0.20cm}{\begin{minipage}[c]{0.65\linewidth}
\raisebox{0.20cm}{\begin{minipage}[c]{0.95\linewidth}
\resizebox{\hsize}{!}{\includegraphics[angle=-90,bb=108 172 460 514,
clip]{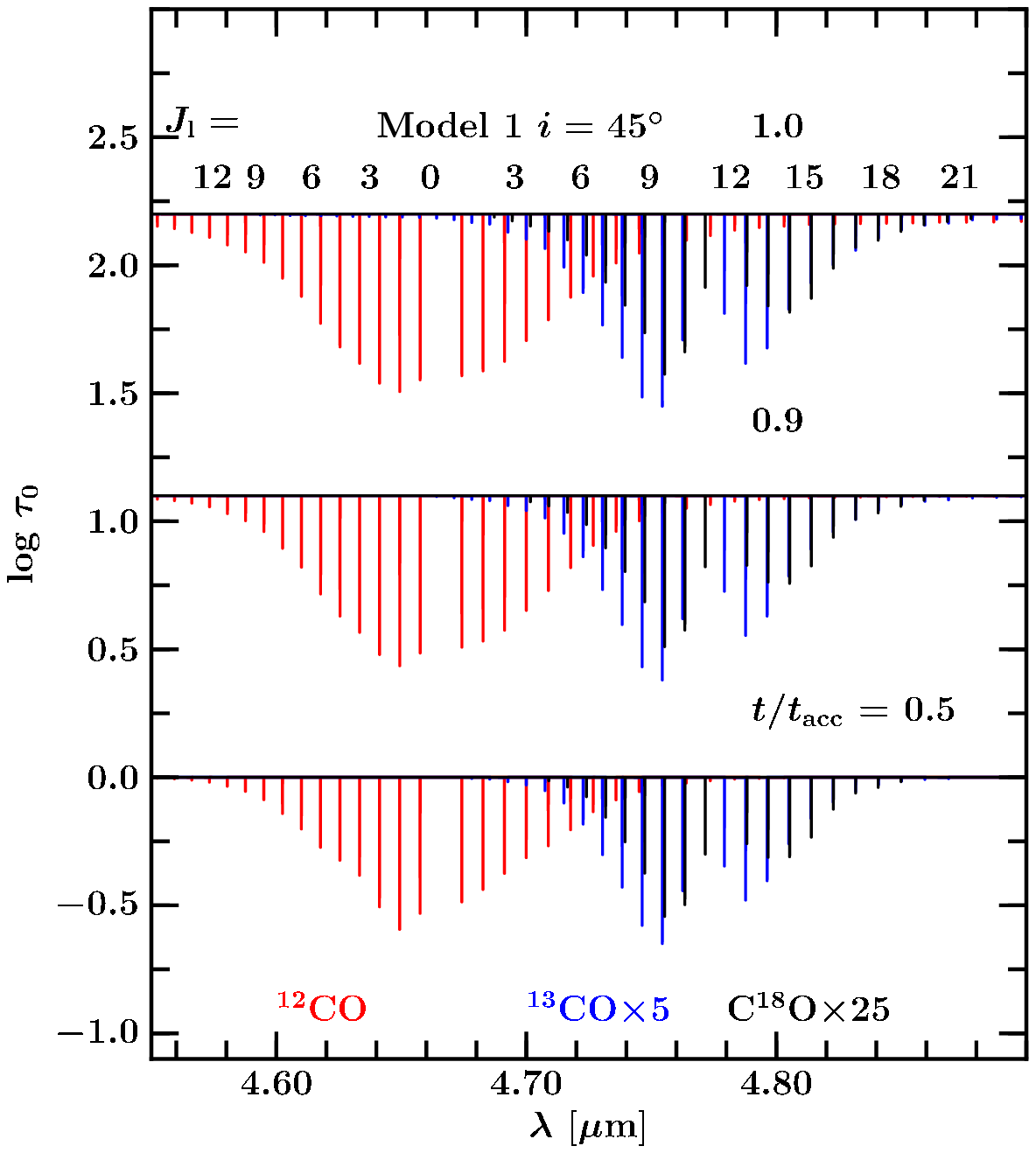}}
\end{minipage}}
\raisebox{0.20cm}{\begin{minipage}[c]{0.95\linewidth}
\resizebox{\hsize}{!}{\includegraphics[angle=-90,bb=108 172 490 514,
clip]{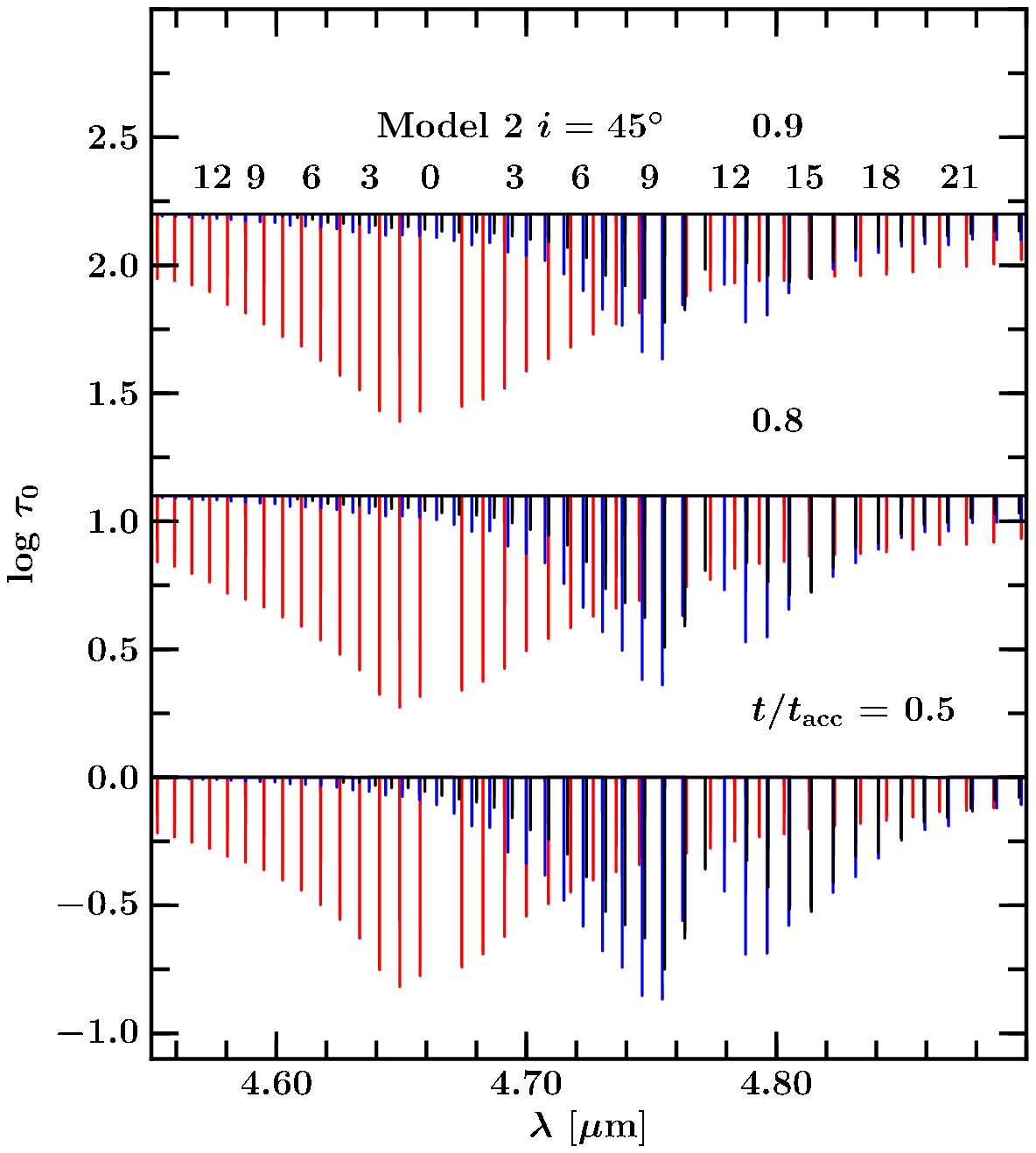}}
\end{minipage}}
\caption{Evolution of the \mco\ (red), \tco\ (blue) and \ceo\ (black) absorption lines for Model 1 at $i = 45^{\circ}$.  The evolution of the absorption lines for Model 2 is shown at
the bottom panel.  The lower rotational levels are indicated on top of the first spectrum.  All of the \tco\ lines are multiplied by 5 and \ceo\ lines by 25.}
\label{fig:nirabs}
\end{figure}

\subsubsection{NIR spectra}

The NIR absorptions are rendered with RADLite for $t/t_{\rm acc} \ge 0.5$
which starts when $M_{\rm env} \sim M_{\star}$ (Stage 0/I boundary) in order to
have a strong 4.7$\mu$m continuum.  Figure~\ref{fig:nirabs} show the P and R
branches of the CO isotopologs for Model 1 and 2 which show the greatest
difference.  In general, as shown in Fig.~\ref{fig:nirabs}, the higher-$J$
absorptions are stronger as the system evolves into Stage II due to
the central luminosity.  For Model 1, at $t/t_{\rm acc} = 0.5$, the
\mco\ absorptions can be seen up to $J = 16$, and up to $J = 9$ for \tco\ and 7
for \ceo.  The number of absorption lines that are above $\tau_{\nu} = 3\times
10^{-3}$ (the 3$\sigma$ observational limit with instruments such as VLT-CRIRES)
doubles at the end of the accretion phase and the highest observable $J$ shifts
from 16 to $>30$ for \mco\ and up to $\sim$ 15--20 for the isotopologs.

At early times, a similar number of lines are found for Model 3 but their
optical depths are lower due to the difference in the inner envelope structure. 
On the other hand, the number of lines in Model 2 is much higher than in the
other two models due to the lower continuum optical depth in the inner
envelope.  Thus, the absorbing column starts deeper than in the other two
models, which results in stronger absorption features and an increased number of
detectable high-$J$ lines,  up to $J> 40$ for \mco, $J\sim 25$ for \tco\ and $J
\sim 15$ for \ceo\ (Fig.~\ref{fig:nirabs}).  Thus, in principle the appearance
of the NIR spectrum could be a sensitive probe of the inner envelope structure.
In practice, the $^{12}$CO absorption will be affected by outflows and winds so
the $^{13}$CO and C$^{18}$O isotopologs are
most useful for this purpose.

\subsection{Rotational temperatures}\label{sec:nirtrot}

A directly related observable is the rotational temperature derived
from a Boltzmann diagram.  Thus, the RADLite NIR spectra need to be
converted into column densities.  For this conversion, we use a standard curve-of-growth analysis \citep{spitzer89} with $b = 0.8$ km s$^{-1}$ and oscillator strengths
derived from the Einstein $A_{\rm ul}$ coefficients.  By using the
curve-of-growth method, it is assumed that the spectral lines are
unresolved.

An example of a Boltzmann diagram is shown in Fig.~\ref{fig:rotdiag1}
together with a two-temperature fit, as commonly done in observations
\citep{mitchell90,smith09,herczeg11}.  The rotational temperature of
the cold component is obtained from fitting the P(1) to P(4) lines
($E_{\mathrm{l}} < 40$ K).  Lines higher than P(5) ($E_{\mathrm{l}} >
40$ K) are used to obtain the temperature of the warm component.
Model 2 at $t/t_{\rm acc} = 0.78$ is used since it clearly shows the
break between the two temperatures.

How do the two temperatures evolve with time?  As Table~\ref{tbl:trotcompare}
shows, the cold component is between 20 and 35 K with no significant evolution,
except in Model 2, as the disk is building up.  The warm component, however,
increases with inclination and time.  The derived warm temperature component
ranges from $< 100$ up to $\sim 520$ K and traces the warming up of material in
the inner region as the envelope dissipates.  From the simulations, a $> 400$ K
warm component is a signature of an evolved system.

\begin{table*}
\centering
\caption{NIR rotational temperatures derived from $\tau_0 > 10^{-3}$ lines for
$i = 45^{\circ}$ and 75$^{\circ}$.}
\label{tbl:trotcompare}
\begin{tabular}{ l l | c c | c c | c c  }
\hline
\hline
& & \multicolumn{2}{c |} { $^{12}$CO } & \multicolumn{2}{|c|}{$^{13}$CO }&
\multicolumn{2}{|c}{C$^{18}$O} \\
$t/t_{\mathrm{acc}}$ & $i$ & $T_{\mathrm{cold}}$ & $T_{\mathrm{warm}}$
& $T_{\mathrm{cold}}$ &
$T_{\mathrm{warm}}$ & $T_{\mathrm{cold}}$ & $T_{\mathrm{warm}}$ \\
 & [$^{\circ}$] & [K] & [K] & [K] & [K] & [K] & [K] \\
\hline
\multicolumn{8}{c}{Model 1} \\
\hline
0.50 & 45 & $21 \pm 9$ & $80 \pm 9$ & $18 \pm 11$ & $49 \pm 18$ & $19 \pm 13$ & ... \\
 & 75 &  $20 \pm 9$ & $66 \pm 8$ & $17 \pm 10$ & $44 \pm 10$ & $18 \pm 10 $ & ... \\
0.76 & 45 & $26 \pm 10$ & $142 \pm 21$ & $19 \pm 11$ & $72 \pm 20$ & $19 \pm 14$ & ... \\
 & 75 &  $25 \pm 12$ & $260 \pm 67$ & $25 \pm 12$ & $273 \pm 69$ & $19 \pm 13 $ & ... \\
0.96 & 45 & $28 \pm 10$ & $201\pm 36$ & $28 \pm 13$ & $207 \pm 46$ & $21 \pm 16$ & ... \\
 & 75 &  $25 \pm 10$ & $271 \pm 61$ & $25 \pm 13$ & $301 \pm 80$ & $20 \pm 13 $ & ... \\
 \hline
\multicolumn{8}{c}{Model 2} \\
\hline
0.50 & 45 & $34 \pm 20$ & $245 \pm 61$ & $34 \pm 25$ & $253\pm61$ & $29\pm23$ & $183 \pm 51$ \\
 & 75 &  $31 \pm 20$ & $293 \pm 85$ & $31 \pm 20$ & $302 \pm 86$ & $27 \pm 19 $ & $209 \pm 67$ \\
0.78 & 45 & $31 \pm 20$ & $327 \pm 100$ & $31 \pm 20$ & $340 \pm 110$ & $28\pm 19$ & $264 \pm 89$ \\
 & 75 &  $20 \pm 11$ & $489 \pm 210$ & $19\pm11$ & $481\pm192$ & $18\pm11$ & $423\pm230$ \\
0.94 & 45 & $28 \pm 18$ & $402 \pm 156$ & $28 \pm 18$ & $410\pm152$ & $26 \pm 17$ & $333\pm139$ \\
 & 75 &  $18 \pm 11$ & $544\pm255$ & $17 \pm 10$ & $527 \pm 223$ & $17 \pm 10$ & $495 \pm 308$ \\
 \hline
\multicolumn{8}{c}{Model 3} \\
\hline
0.50 & 45 & $21 \pm 6$ & $193 \pm 37$ & $16 \pm 8$ & $42 \pm 21$ & $19 \pm 9$ & ... \\
 & 75 &  $16 \pm 7$ & $37 \pm 7$ & $16 \pm 10$ & ... & $16 \pm 11 $ & ... \\
0.76 & 45 & $23 \pm 9$ & $287 \pm 80$ & $22 \pm 9$ & $304 \pm 85$ & $17 \pm 11$ & ... \\
 & 75 &  $22 \pm 8$ & $289 \pm 81$ & $21 \pm 8$ & $302 \pm 83$ & $17 \pm 10$ & ... \\
0.96 & 45 & $22 \pm 9$ & $366 \pm 128$ & $21 \pm 8$ & $381 \pm 129$ & $17 \pm 10$ & ... \\
 & 75 &  $18 \pm 8$ & $361 \pm 121$ & $18\pm7$ & $371 \pm 119$ & $16 \pm 9$ & ... \\
\hline
%  \hline
 \end{tabular}
\end{table*}

\begin{figure}
 \centering
\raisebox{0.20cm}{\begin{minipage}[c]{1.00\linewidth}
\resizebox{\hsize}{!}{\includegraphics[angle=0,bb=30 15 550 415, clip]{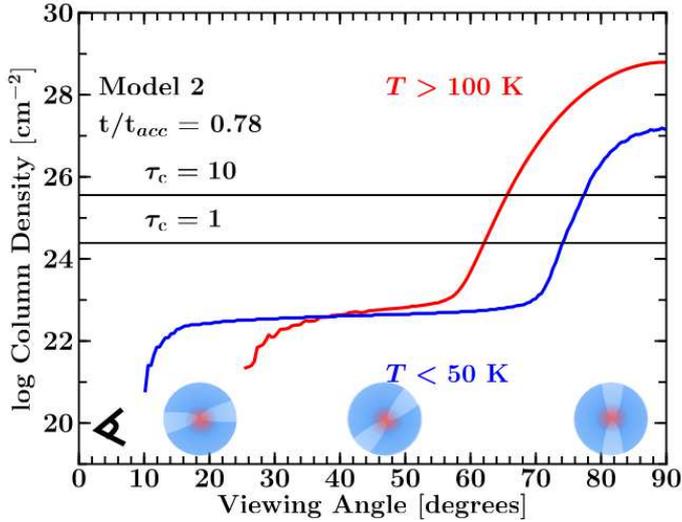}}
\end{minipage}} 
\caption{Integrated gas column density along the different viewing angles
for Model 2.  The red and blues lines indicate the warm ($T > 100$ K) and cold
gas ($T < 50$ K), respectively.  The approximate continuum optical depths,
$\tau_{\rm c}=1$ and 10, are also shown to indicate the amount material that is
missed by properly solving the radiative transfer equation. }
 \label{fig:h2col1}
\end{figure}

What is the origin of the two temperatures? This can be easily shown by plotting
the cold ($<$50 K) and warm ($>$100 K) H$_2$ column along the line of sight at
various inclinations (Fig.~\ref{fig:h2col1}).  The particular model and time
were chosen as an example; the general picture is the same independent of
evolutionary model, but the warm component shifts to lower viewing angle (left)
for earlier times.  The figure indicates that the warm material can be probed
through $i > 25^{\circ}$ viewing angles.  The cold component probes the outer
envelope where the gas temperature hardly changes as the system evolves. The
warm component reflects the line of sight to the center through the warm inner
envelope (the so-called `hot core' ) or the disk surface where the gas is $>
100$ K in all of our models. The cumulative gas column density indicates that
the cold material is mostly absorbed at $R > 100$ AU while the warmer gas
component is probing the inner few AU ($i > 60^{\circ}$) up to $\sim$20 AU ($i
\sim 45^{\circ}$).  The excitation temperature of the warm component reflects
the density structure in the inner few AU.  With the lower densities and warmer
temperatures in Model 2, which decreases the effective continuum optical depth,
higher warm temperatures can be observed as the absorption probes deeper to
smaller radii than the other two models.

%%%%%%%%%%%%%%%%%%%%%%%%%%%%%%%%%%%%%%%%%%%%%%%%%%%%%%%%%%%%%%%%%%%%%%%%%%%%
%%%%%%%%%%%%%%%%%%%%%%%%%%%%%%%%%%%%%%%%%%%%%%%%%%%%%%%%%%%%%%%%%%%%%%%%%%%%

\section{Discussion}\label{sec:dis}

We have presented the evolution of CO molecular lines based on the standard
picture of a prestellar core collapsing to form a star and 2D circumstellar
disk. The main aim of this work is to study the evolution of observables that are
derived from the spectra such as rotational temperatures, line profiles, and
velocity fields. Specifically, signatures of disk formation during the embedded
stage that can be obtained with the most commonly used molecule, CO, are
investigated.

\subsection{Detecting disk signatures in the embedded phase}

How can one determine whether disks have formed in the earliest Stage
0?  As discussed in Section~\ref{sec:diskcon}, it is difficult to isolate
an embedded disk component in single-dish observations.  The disk contribution
to the observed flux within $> 9 \arcsec$ beams is only significant ($>$60\%)
for the \tco\ and \ceo\ $J_{\rm u} > 9$ lines at $t/t_{\rm acc} > 0.75$. 
However, the absolute fluxes are too low to be detected.  Thus, spatially
resolved observations at sub-arcsec scale are needed.

\begin{figure}
 \centering
\raisebox{0.20cm}{\begin{minipage}[c]{1.00\linewidth}
\resizebox{\hsize}{!}{\includegraphics[angle=0,bb=105 238 435 533,
clip]{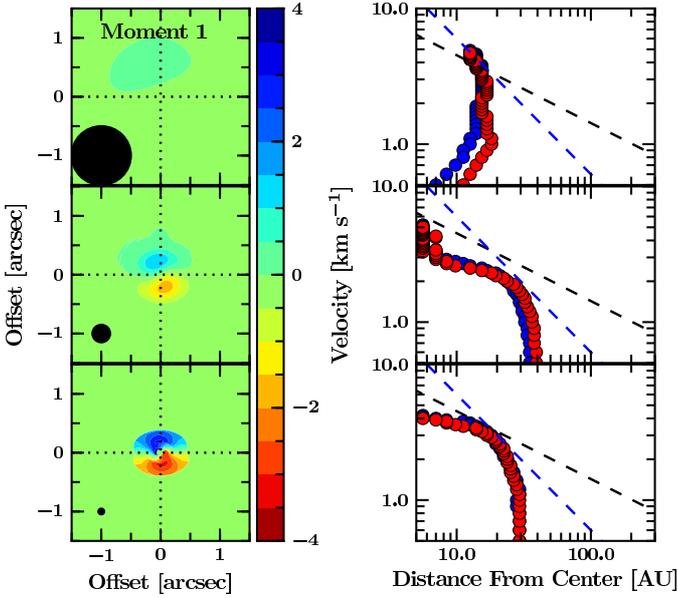}}
\end{minipage}}
\caption{\ceo\ 3--2 moment-one (left) and spectroastrometry (right) for
  Model 1 at the end of Stage 0 convolved to a 1$\arcsec$ (top),
  0.3$\arcsec$ (middle) and 0.1$\arcsec$ (bottom) as shown by the
  solid circle.  The Keplerian structure at this stage extends up to 20 AU
(0.14$\arcsec$ radius).  The black dashed line in the right panel shows the
$v\propto R^{-0.5}$ and the blue dashed line indicates the $v\propto R^{-1}$
relation.}
 \label{fig:stage0}
\end{figure}

What are the signatures of an embedded rotationally supported disk on
subarcsecond scales?  We have analyzed some of the simulated images during the
Stage 0 phase at a resolution of 0.1$\arcsec$ as can readily be observed with
ALMA.  Figure~\ref{fig:stage0} shows the \ceo\ 3--2 moment-one map during the
Stage 0 phase of Model 1 with a disk extending up to 20 AU at a resolution of 1,
0.3 and 0.1$\arcsec$.  Velocity gradients are present in all of the moment-one
maps although it is weaker within a 1$\arcsec$ beam.  The right panel of
the figure shows the spectroastrometry (Fig.~\ref{fig:pv2}) within the
different resolutions.  The velocity gradient within a 0.5--1$\arcsec$ beam
exhibits a $v \propto R^{-1}$ relation since the inner envelope dominates the
emission.  Within the smaller beams, the velocity gradient is resolved into two
components: $v \propto R^{-1}$ (blue dashed line) due to envelope and $v \propto
R^{-0.5}$ (black dashed line) due to the stable disk.  A  $v \propto R^{-0.5}$
can be seen when the disk is marginally resolved, however, the derived stellar
mass is far below the real stellar mass.  It is very important that the
disk is fully resolved in order to derive the stellar mass.  With the \ceo\
lines, a resolved embedded disk in Stage 0 would have the following features:
\begin{itemize}
 \item Elongated moment-zero map.
 \item A transition between an infalling envelope to a rotating disk: a skewed
moment-one map.
 \item The Keplerian structure exhibits a clear $ v\propto R^{-0.5}$
(Fig.~\ref{fig:pv2} and bottom of Fig.~\ref{fig:stage0} ) velocity gradient
perpendicular to the outflow.
\end{itemize}

It is possible to search for rotationally supported
disks in Stage 0 sources with characteristics of Model 1 with the full ALMA
array in less than 30 minutes for the C$^{18}$O 2--1 transition or up to 4 hours
for the 6--5 transition.  For Models 2 and 3, the disks have grown up to $\sim
50$ AU near the end of Stage 0 and up to twice more massive, thus it will take
less than 2 hours to observe them with ALMA in the 6--5 transition and less than
30 minutes in the 2--1 and 3--2 transitions. Therefore, ALMA will allow us, for
the first time, to test whether rotationally supported disks have grown beyond
10 AU by the end of the Stage 0 phase.

The question of when rotationally supported disks form is crucial to
the physical and chemical evolution of accretion disks.  In our evolutionary
models, the disks have radii up to 50 AU at the end of Stage 0 phase and can be
as small as 20 AU.  \citet{dapp10} and \citet{dapp12} numerically showed that
disks do not grow beyond 10 AU during the Stage 0 phase in the presence of
magnetic fields due to the magnetic breaking problem \citep[see also][]{zyli11,joos12}. 
This is consistent with the lack of observed rotationally supported disks toward
Stage 0 objects so far \citep{brinch09,maury10}.  On the other hand, there is a growing body of observational evidence for rotating disks in Stage I YSOs \citep{brinch07,lommen08,prosac09,takakuwa12}.  
The size of the rotationally supported disk depends on the initial parameters of
the collapsing envelope.  In our models, the sizes of Models 2 and 3 are
consistent with the observed Keplerian velocity structure in Stage I sources
\citep{prosac09}.  

An additional caveat is that our models lead to a Keplerian disk with a
flattened inner envelope on similar scales as the stable disk.  
Hydrodynamical simulations, on the other hand, show a stable disk
embedded in a much larger rotating flattened structure and with more turbulent
structure \citep{brinch08, kratter10}.  Furthermore, the 3D simulations by
\citet{hal11} suggest that 50\% of the embedded disk is sub-Keplerian due to
interaction with the envelope.  The predicted moment 1 maps and
spectroastrometry with the current models should be performed for these
numerical simulations for comparison with ALMA observations.

\subsection{Probing the temperature structure}

The derived rotational temperatures from the submm and FIR lines
($J_{\rm u}$ = 1--10) within $> 9\arcsec$ beams do not evolve with
time, in contrast with the continuum SED, and largely trace the outer
envelope (Section~\ref{sec:trots}).  Even though the disk structure is hotter
with time, the emission that comes from within that region is much smaller than
the beam.  The mass weighted temperature of the system which contributes to the
overall emission is constant with time.  The rotational temperatures also cannot
differentiate between the three different evolutionary models.  In addition, the
peak of the CO SLED is constant throughout the evolution at $J_{\rm u}=4$
for the \tco\ and \ceo.  The higher observed excitation temperatures for \mco\
and \tco\ in a wide range of low-mass protostars from submm and FIR data
point to other physical processes that are present in the system such as UV
heating and C-shock components that affect the $J\ge 5$ lines
\citep[][]{visser12,yildiz12}.

The NIR absorption lines toward embedded YSOs are complementary to
the submm/FIR rotational lines.  Both techniques probe the bulk of the cold
envelope, but the NIR lines also probe the warm component because the lines
start absorbing at the $\tau \sim 1$ surface deep inside the inner envelope.
Consequently, the cold component present in the NIR lines should be similar
to the component observed in the submm. Indeed, the typical rotational
temperatures of 20--30 K found in the NIR models for \ceo\ are
comparable to the values of 30--40 K found from the submm simulations.

The warm up process is accessible through NIR molecular absorption lines. 
In Section~\ref{sec:nirtrot}, it is found that the warm component of the
Boltzmann diagram evolves with time.  The derived warm temperatures have a large
spread depending on inclination and density structure in the inner few AU.  A
higher inner envelope and disk density correspond to a $\sim 100$ K warm
component, whereas a more diffuse inner envelope has a warmer temperature
that can be up to 500--600 K.  Thus, as long as the inclination is known, the
temperature of the warm component may give us a clue on the density
structure of the inner few AU where the $>100$ K gas resides.

The predicted cold and warm temperatures can be compared with those
found in NIR data toward Stage I low-mass YSOs.  For the cold component, a
temperature of 15 K has been found for L1489 from \tco\ data \citep{boogert02},
and 10--20 K for Reipurth 50, Oph IRS43 and Oph IRS 63 from \tco\ and \ceo\ data
\citep{smith09,smith10}, consistent with our envelope models.  For the warm
component, values ranging from 100 to 250 K are found for these sources from the
isotopolog data.  These warm rotational temperatures are on the low side of the
model values in Table~\ref{tbl:trotcompare} for the later evolutionary stages,
and may indicate either a compact envelope (high inner envelope densities)
and/or high inclination (which effectively limits the absorption lines to probe
the inner envelope component).  Independent information on the inclination is
needed to further test the physical structure of those sources.

%%%%%%%%%%%%%%%%%%%%%%%%%%%%%%%%%%%%%%%%%%%%%%%%%%%%%%%%%%%%%%%%%%%%%%%%%%%%%%%
%%%%%%%%%%%%%%%%%%%%%%%%%%%%%%%%%%%%%%%%%%%%%%%%%%%%%%%%%%%%%%%%%%%%%%%%%%%%%%%

%\clearpage

\section{Summary and conclusions}\label{sec:sum}

Two-dimensional self-consistent envelope collapse and disk formation models have
been used to study the evolution of molecular lines during the embedded phase of
star formation.  The non-LTE molecular excitation has been computed with an
escape probability method and spectral cubes have been simulated for both the
FIR and NIR regimes.  The gas temperature is taken to be well coupled to
the dust temperature as obtained through a continuum full radiative transfer
code.  With the availability of spectrally resolved data from single dish
submillimeter telescopes, the {\it Herschel} Space Observatory, VLT CRIRES and
ALMA, it is now possible to compare the predicted collapse dynamics with
observations. We have focused the analysis on the \tco\ and the \ceo\ lines
since \mco\ lines are dominated by outflows and UV heating. The main
conclusions are as follows:

\begin{itemize}

\item Spectrally resolved molecular lines are important in comparing
  theoretical models of star formation with observations and
  to distinguish the different physical components.  The collapsing
  rotating envelope can readily be studied through the \ceo\ lines.
  Their FWHM probes the collapse dynamics, especially for the higher-$J$
($J_{\rm u} \ge 6$) lines where up to 50\% of the line broadening can be due to
infall.

\item The derived rotational temperatures and SLED from submm and   FIR pure
rotational lines are found to be independent of   evolution and do not probe the
warm up process, in contrast to the continuum SED.  The predicted rotational
  temperatures are consistent with observations for \ceo\ for a large
  sample of low-mass protostars.

\item The predicted \mco\ and \tco\ rotational temperatures and high-$J$
  fluxes are lower than those found in observations of a wide variety
  of low-mass sources \citep{goicoechea12,yildiz12,yildiz13a}. 
This indicates the presence of additional physical processes that heat the gas
such as shocks and UV heating of the cavity walls \citep{visser12}.

\item The NIR absorption lines are complementary to the
  FIR/submm lines since they probe both the cold outer envelope and
  the warm up process.  Values obtained for the cold component are
  consistent with the observational data and models. For the warm
  component, the observed values are generally on the low side
  compared with the model results. The high-$J$ NIR lines are
  strongly affected by radiative transfer effects, which depend on the
  physical structure of the inner few AU and thus form a unique probe
  of that region.

\item The simulations indicate that an embedded disk in both Stage 0 ($M_{\rm
env} > M_{\star}$) and I ($M_{\rm env} < M_{\star}$ but $M_{\rm disk} < M_{\rm
env}$) does not contribute significantly ($< 50$\%) to the emergent  $J_{\rm u}
< 8$ lines within $> 9 \arcsec$ beams. Higher-$J$ isotopolog lines have a higher
contribution but are generally too weak to observe.  The disk contribution
is significantly higher within 1$\arcsec$ beam and the evolution within such a
beam indicate rotationally supported disks should be detectable in Stage I
phase consistent with observations \citep{prosac09}.

\item Embedded disks during the Stage I phase are generally large
  enough to be detected with current interferometric instruments with
  $1''$ resolution, consistent with observations.  On the other hand,
  high signal-to-noise ALMA data at $\sim 0.1''$ resolution are needed in order
  to find signatures of embedded disks during the Stage 0 phase. A
  careful analysis is needed to disentangle the disk from the envelope.

\item We have shown that the rotationally dominated disk can be disentangled
from the collapsing rotating envelope with high signal-to-noise and high 
spectral resolution interferometric observations (Section~\ref{sec:evolvels}).  
The spectroastrometry with ALMA by plotting the velocity as function of peak
positions can reveal the size of the Keplerian disk.  The \tco\ lines within
interferometric observations may still be contaminated by the infalling rotating
envelope.  Thus, more optically thin tracers are required.

\end{itemize}

The three different collapse models studied here differ mostly in their physical
structure and velocity fields on 10--500 AU scales and illustrate the range of
values that are likely to be encountered in observational studies of embedded
YSOs.  It is clear that the combination of spatially and spectrally resolved
molecular line observations by ALMA and at NIR are crucial in determining
the dynamical processes in the innermost regions during the early stages of
star formation.  A comparison between the standard picture presented here or
hydrodynamical simulations and molecular line observations of Stage 0 YSOs will
further test the theoretical picture of star and planet formation.

%%%%%%%%%%%%%%%%%%%%%%%%%%%%%%%%%%%%%%%%%%%%%%%%%%%%%%%%%%%%%%%%%%%%%%%%%%%%%%
%%%%%%%%%%%%%%%%%%%%%%%%%%%%%%%%%%%%%%%%%%%%%%%%%%%%%%%%%%%%%%%%%%%%%%%%%%%%%%

\section*{Acknowledgements}

We would like to thank Steve Doty for stimulating discussions on continuum and
line radiative transfer and for allowing us to use his chemistry code.  We are
also grateful to Kees Dullemond for providing RADMC (and RADMC3D) and to
Klaus Pontoppidan for RADLite. { We thank the anonymous referee for 
the constructive comments, which have improved this paper}.  This work is 
supported by the Netherlands Research School for Astronomy (NOVA) and by the 
Space Research Organization
Netherlands (SRON).  Astrochemistry in Leiden is supported by the
Netherlands Research School for Astronomy (NOVA), by a Spinoza grant and grant
614.001.008 from the Netherlands Organisation for Scientific Research (NWO), and
by the European Community's Seventh Framework Programme FP7/2007–2013 under
grant agreement 238258 (LASSIE).  

%%%%%%%%%%%%%%%%%%%%%%%%%%%%%%%%%%%%%%%%%%%%%%%%%%%%%%%%%%%%%%%%%%%%%%%%%%%%%%
%%%%%%%%%%%%%%%%%%%%%%%%%%%%%%%%%%%%%%%%%%%%%%%%%%%%%%%%%%%%%%%%%%%%%%%%%%%%%%

\bibliographystyle{aa}
\bibliography{AA201220885}

%%%%%%%%%%%%%%%%%%%%%%%%%%%%%%%%%%%%%%%%%%%%%%%%%%%%%%%%%%%%%%%%%%%%%%%%%%%%%%
%%%%%%%%%%%%%%%%%%%%%%%%%%%%%%%%%%%%%%%%%%%%%%%%%%%%%%%%%%%%%%%%%%%%%%%%%%%%%%

% \clearpage
\appendix

\Online

\section{Two dimensional RT grid}\label{app:A}

\begin{figure}[h]
\centering
\begin{tabular}{c}
\raisebox{0.20cm}{\begin{minipage}[c]{0.90\linewidth}
\resizebox{\hsize}{!}{\includegraphics[angle=0,bb=35 220 565
572,clip]{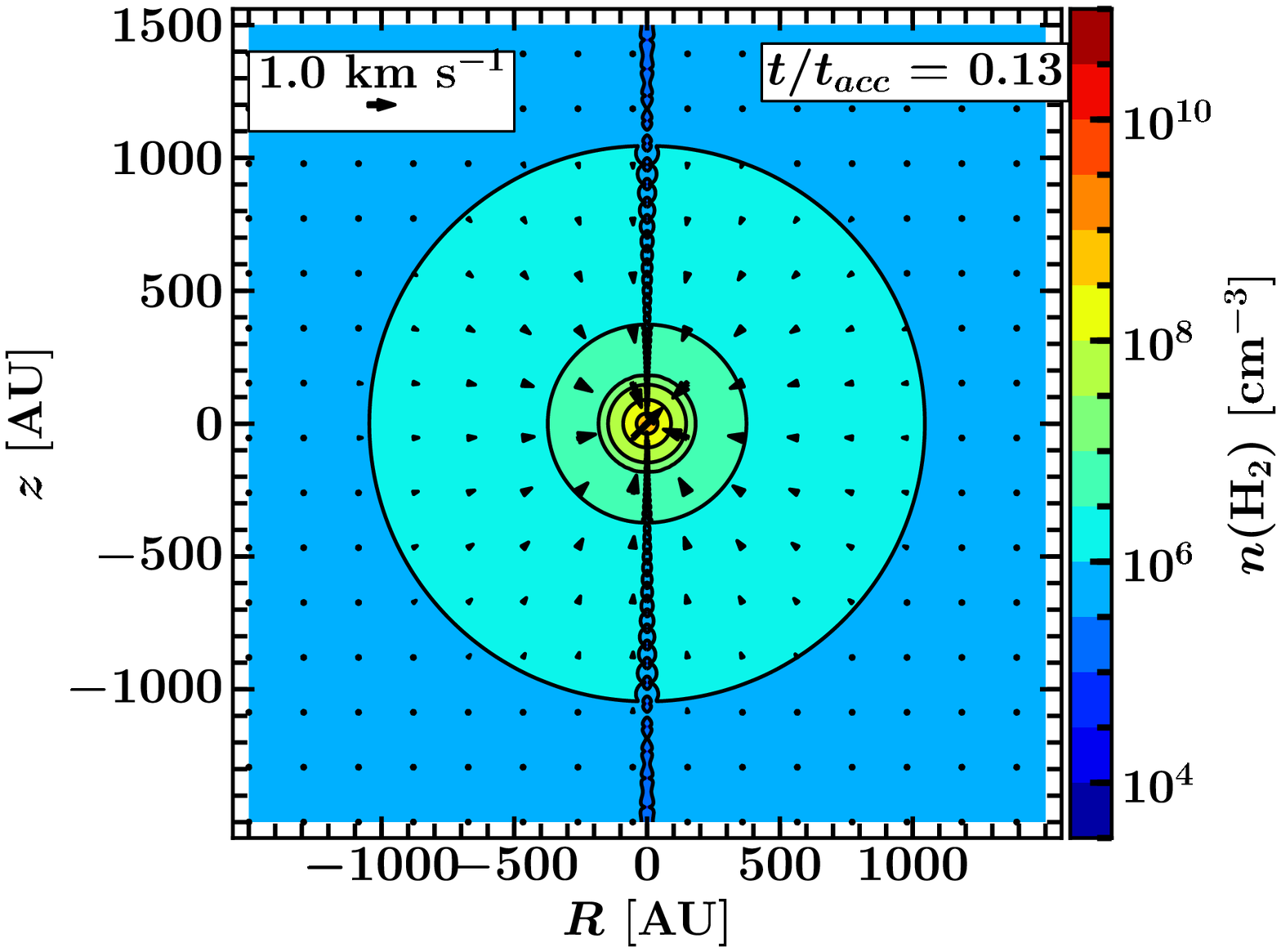}}
\end{minipage}} \\
\raisebox{0.20cm}{\begin{minipage}[c]{0.90\linewidth}
\resizebox{\hsize}{!}{\includegraphics[angle=0,bb=35 175 565
572,clip]{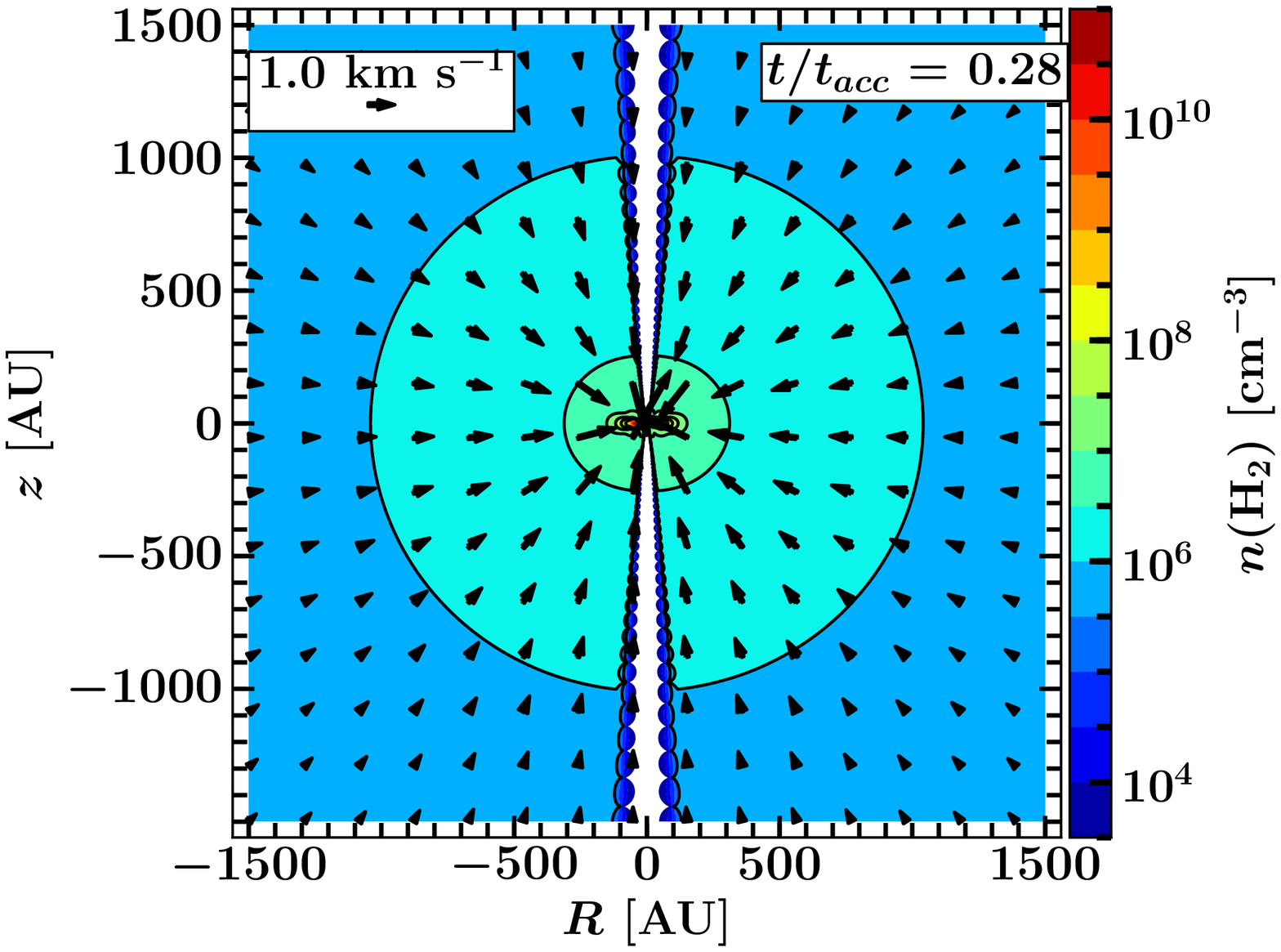}}
\end{minipage}}  
\end{tabular}
\caption{Gas density and velocity field for Model 3 at
  $t/t_{\rm acc} = 0.13$ and 0.28 within 1500 AU.  The density contours start 
from $\log (n/{\rm cm}^{-3}) = 5.5$ and increase by steps of 0.5 up to
  $\log (n/{\rm cm}^{-3}) = 9$.  There are vertical and radial motions
  in the disk, which are not captured in this figure.  }
\label{fig:struct1}
\end{figure}

The grid needs to resolve the steep density and velocity gradients at the
boundaries between adjacent components as shown in Fig.~\ref{fig:struct1}. 
Improper gridding can lead to order-of-magnitude difference in the high-$J$ ($J
\ge 6$) line fluxes. The escape probability code begins with a set of regularly
spaced cells on a logarithmic grid to resolve both small and large scales. Any
cells where the abundance or density at the corners differs by more than a
factor of 5, or where the temperature at the corners differs by more than a
factor of 1.5, are split into smaller cells until the conditions across each
cell are roughly constant. The full non-LTE excitation calculation takes about
five minutes for a typical number of cells of 15\,000 (Stage I) -- 25\,000
(Stage 0). 

\begin{figure}
 \centering
 \raisebox{0.20cm}{\begin{minipage}[c]{1.00\linewidth}
\resizebox{\hsize}{!}{\includegraphics[angle=0,bb = 18 21 575 435, clip ]{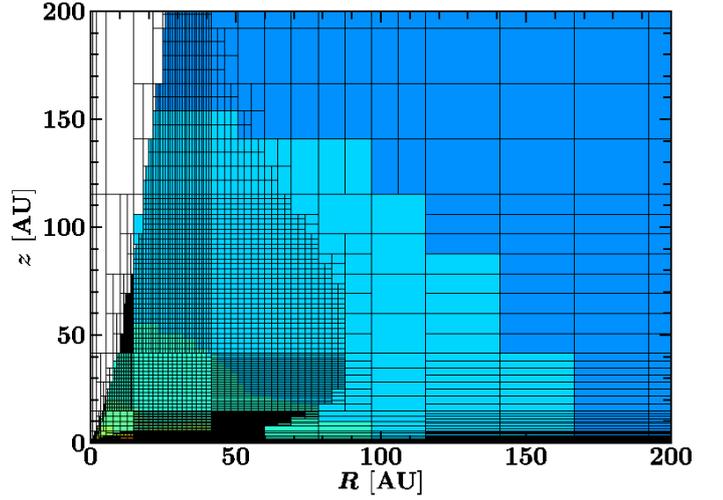}}
\end{minipage}} 
\caption{Example of the gridding used in the Stage 0 phase to resolve the
small extent of the outflow cavity and to resolve the density and velocity
gradient from the disk midplane to the envelope.  The color contours are the
same as in Fig.~\ref{fig:struct1}.  The structure shown is for Model 1 at
$t/t_{\rm acc} = 0.13$.}
 \label{fig:gridexample}
\end{figure}

Figure~\ref{fig:gridexample} presents an example of the gridding in our models. 
Such gridding is most important for early time steps where the outflow cavity
opening angle is small.  The refining ensures that the non-LTE population
calculation converges and high-$J$ emission which comes from the inner
region can escape.

%%%%%%%%%%%%%%%%%%%%%%%%%%%%%%%%%%%%%%%%%%%%%%%%%%%%%%%%%%%%%%%%%%%%%%%%%%%%%%%
%%%%%%%%%%%%%%%%%%%%%%%%%%%%%%%%%%%%%%%%%%%%%%%%%%%%%%%%%%%%%%%%%%%%%%%%%%%%%%%

\section{FIR and submm lines}\label{app:B}

\begin{figure}
 \centering
 \raisebox{0.20cm}{\begin{minipage}[c]{1.00\linewidth}
\resizebox{\hsize}{!}{\includegraphics[angle=0,bb=5 183 537 589,
clip]{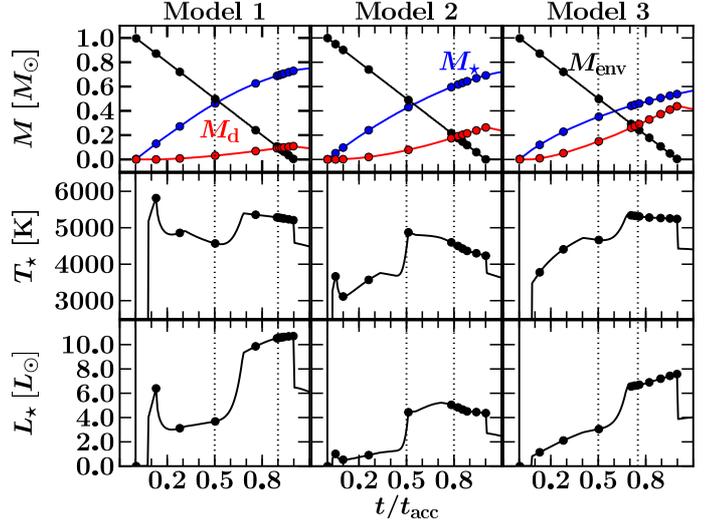}}
\end{minipage}}
\caption{ Evolution of the envelope, disk and stellar mass (top),
  effective stellar temperature (middle) and stellar luminosity
  (bottom) for the three different models as function of time (in
  units of $t_{\mathrm{acc}}$).  The solid circles show the time steps
  (roughly around $t/t_{\rm acc} \sim 10^{-3}, 0.1, 0.5, 0.75, 0.89,
  0.99$) used for rendering the molecular lines.  The adopted time steps
    vary per model in order to cover properly the time when the model enters
Stage 1 ($M_{\rm env} < M_{\star}$) and
when the $M_{\rm d} = M_{\rm env}$ as indicated by the vertical dotted lines.}
 \label{fig:evol1}
\end{figure}

\begin{figure}
 \centering
  \raisebox{0.25cm}{\begin{minipage}[c]{1.00\linewidth}
\resizebox{\hsize}{!}{\includegraphics[angle=0,bb=33 185 561 478, clip]{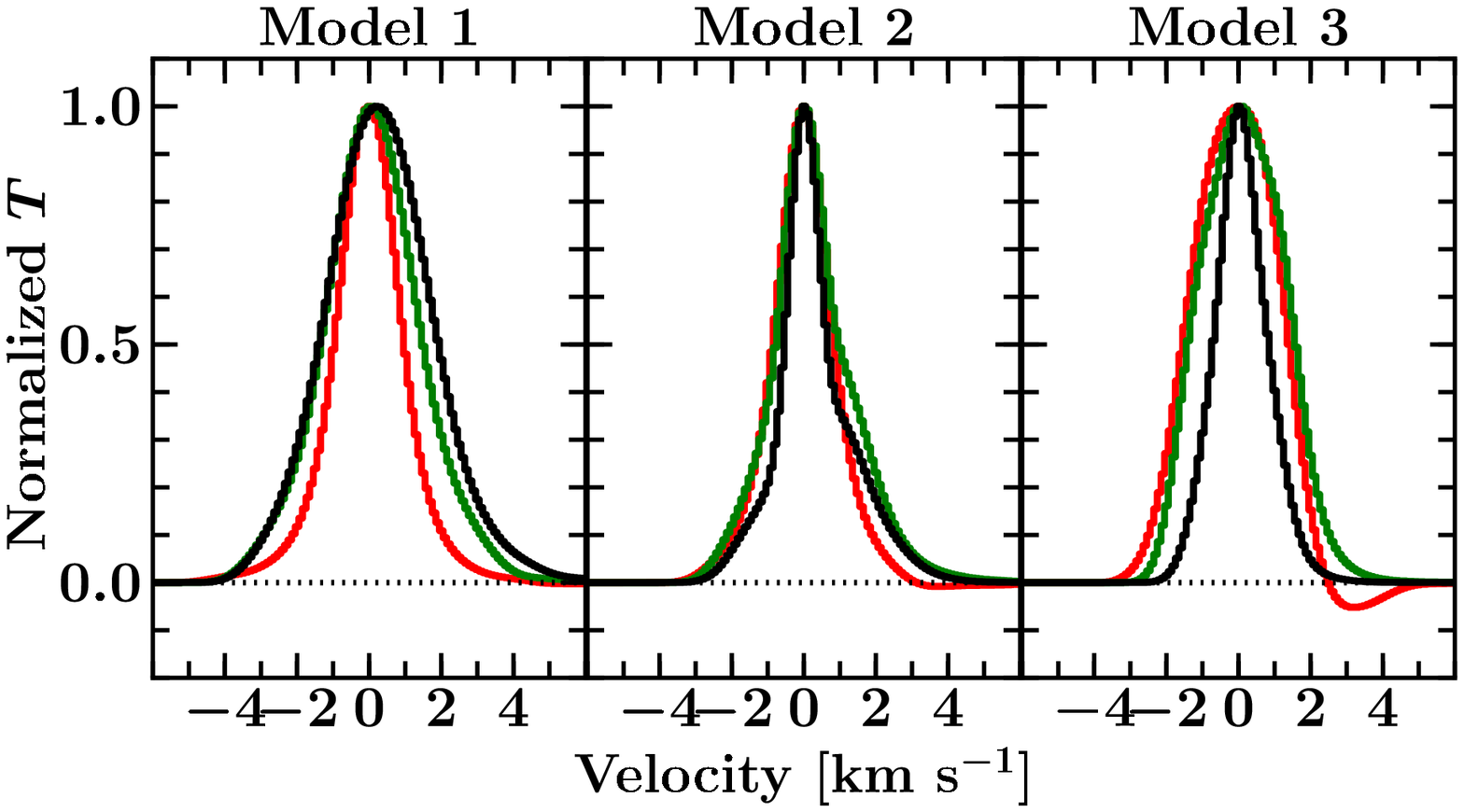}}
\end{minipage}} 
\caption{ \ceo\ $9-8$ line profiles convolved to a 9$\arcsec$ beam for the three models at $t/t_{\mathrm{acc}} = 0.13$
(red), 0.50 (green) and 0.96 (black).}
 \label{fig:allc18o9}
\end{figure}

\begin{figure}
 \centering
  \raisebox{0.25cm}{\begin{minipage}[c]{1.00\linewidth}
\resizebox{\hsize}{!}{\includegraphics[angle=0,bb=35 175 540 595,
clip]{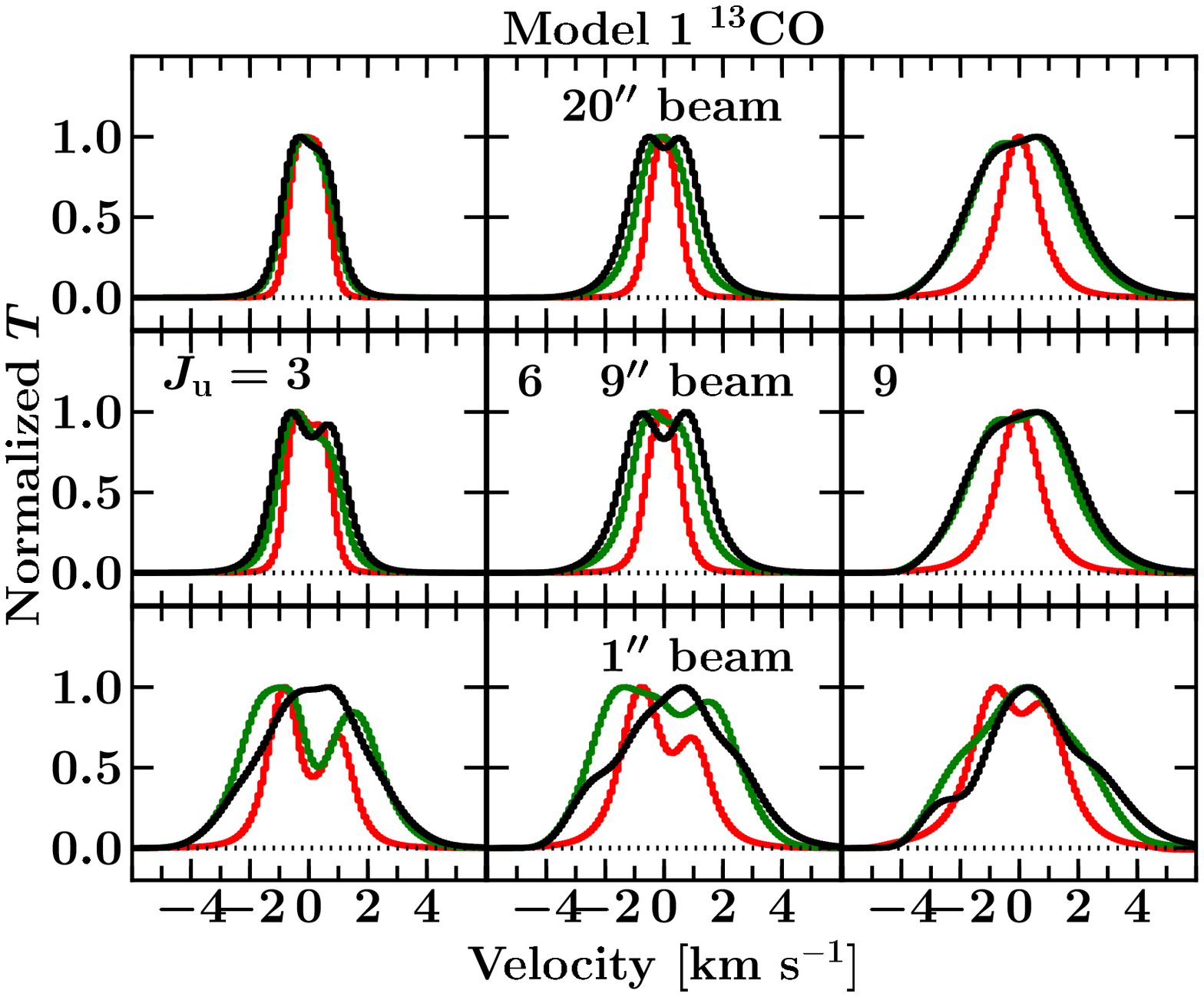}}
\end{minipage}}
\caption{ Normalized \ceo\ and \tco\ line profiles as functions of evolution
for $J_{\mathrm{u}} = 3$, 6 and 9 at $t/t_{\mathrm{acc}} = 0.13$ (red), 0.50
(green) and 0.96 (black) for $i = 5^{\circ}$ orientation for Model 1.  The lines
are convolved to beams of 20$\arcsec$ (top), 9$\arcsec$ (middle) and 1$\arcsec$
(bottom).}
 \label{fig:13colines}
\end{figure}

\begin{figure*}
\begin{tabular}{cc}
 \centering
   \raisebox{0.20cm}{\begin{minipage}[c]{0.5\linewidth}
\resizebox{\hsize}{!}{\includegraphics[angle=0,bb=35 220 540 595, clip]{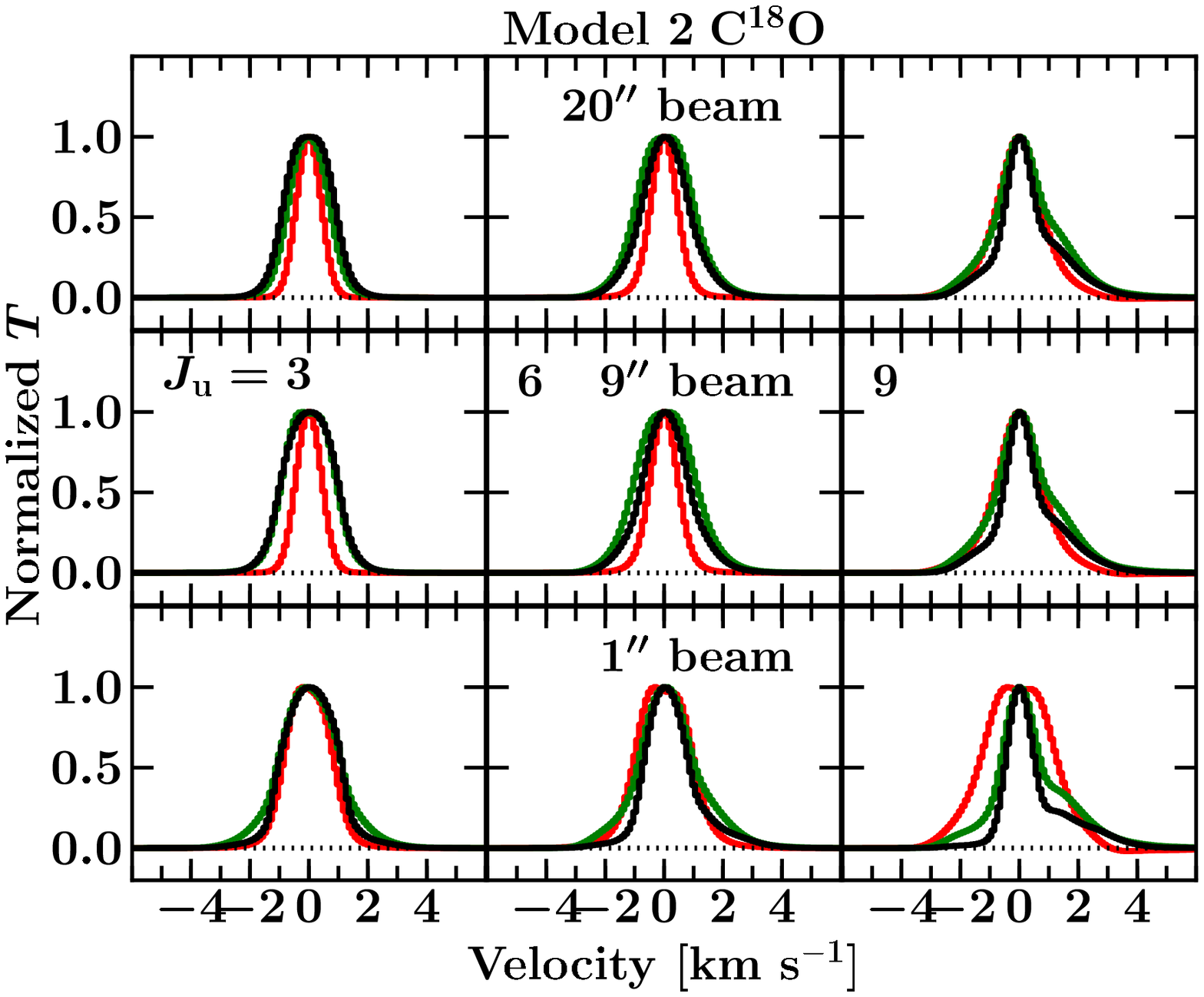}}
\end{minipage}} &
   \raisebox{0.20cm}{\begin{minipage}[c]{0.45\linewidth}
\resizebox{\hsize}{!}{\includegraphics[angle=0,bb=87 220 540 595, clip]{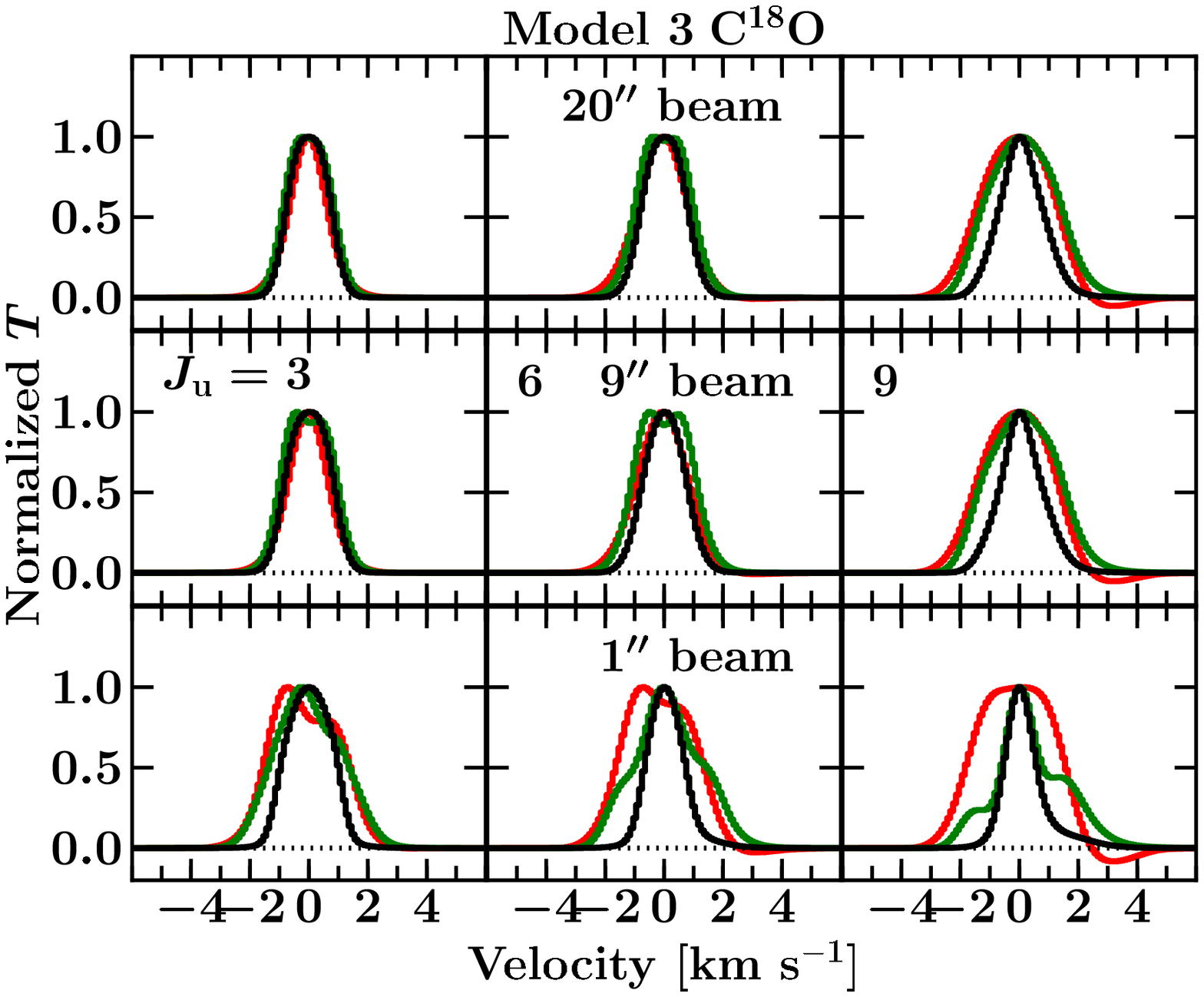}}
\end{minipage}} \\
   \raisebox{0.20cm}{\begin{minipage}[c]{0.5\linewidth}
\resizebox{\hsize}{!}{\includegraphics[angle=0,bb=35 175 540 595, clip]{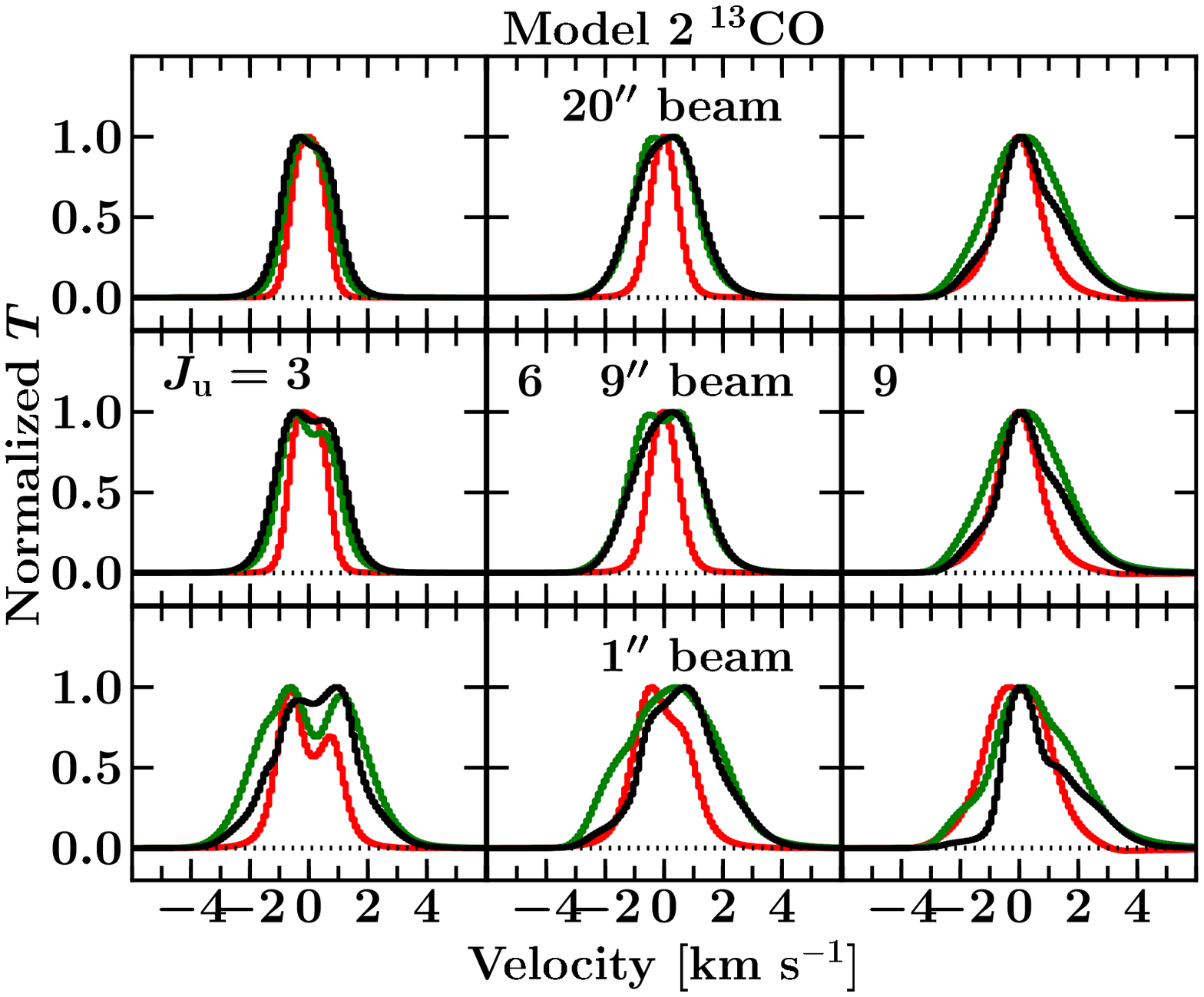}}
\end{minipage}} &
   \raisebox{0.20cm}{\begin{minipage}[c]{0.45\linewidth}
\resizebox{\hsize}{!}{\includegraphics[angle=0,bb=87 175 540 595, clip]{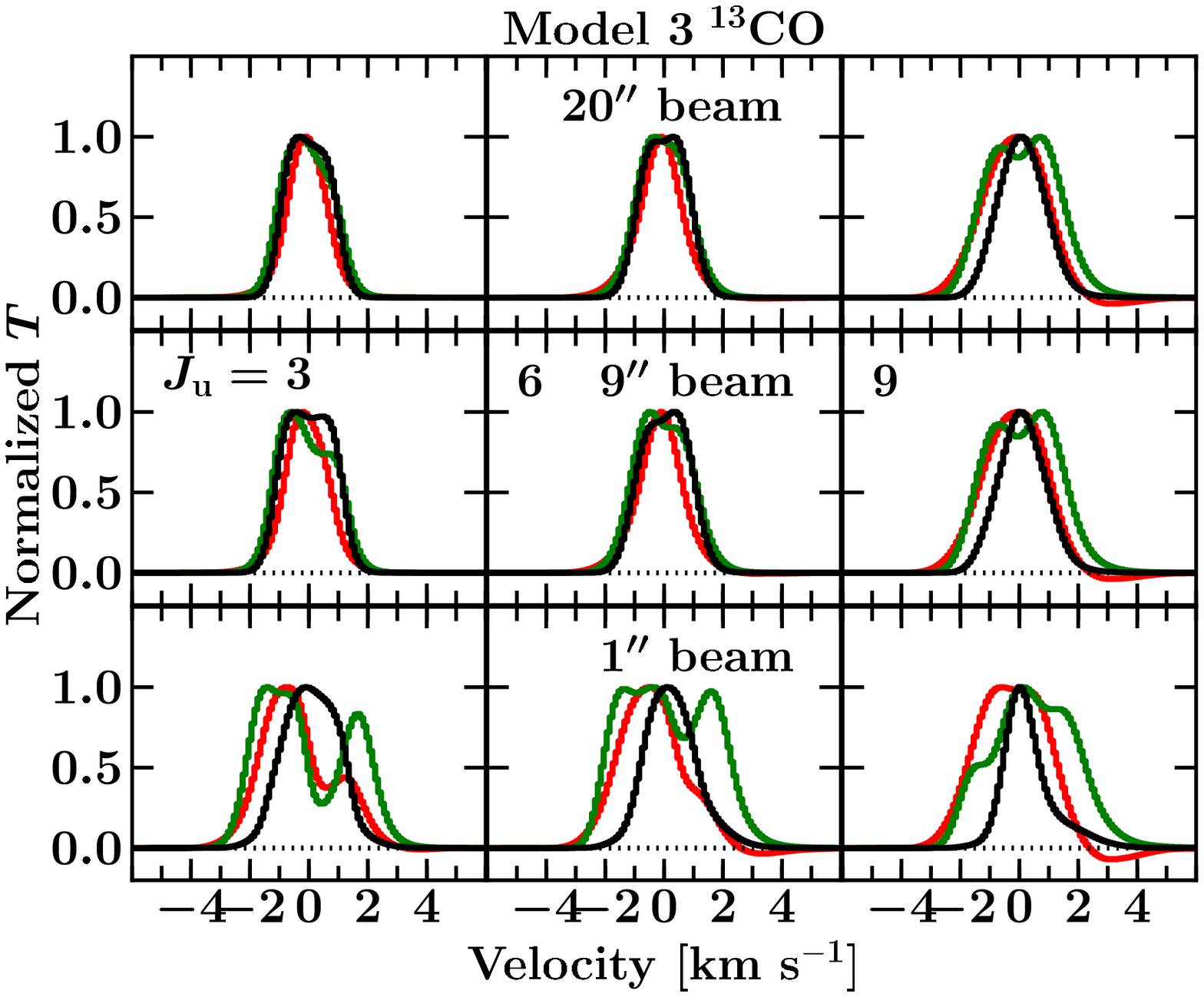}}
\end{minipage}}
\end{tabular}
\caption{ As Figure\ref{fig:c18olines}, but for Models 2 and 3.}
 \label{fig:evollines}
\end{figure*}

\begin{figure}
 \centering
   \raisebox{0.20cm}{\begin{minipage}[c]{1.00\linewidth}
\resizebox{\hsize}{!}{\includegraphics[angle=0,bb=35 180 540 590, clip]{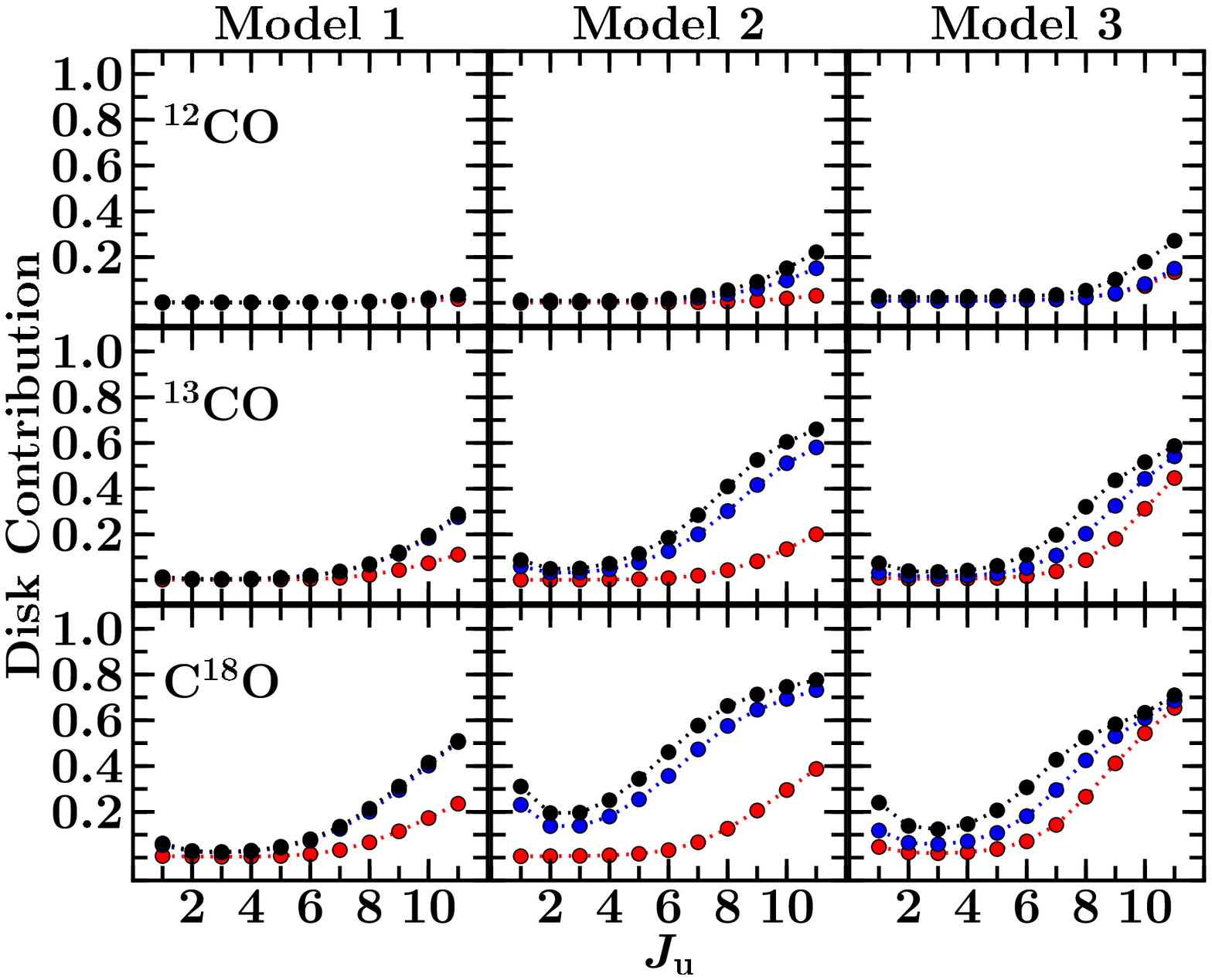}}
\end{minipage}} 
\caption{ Disk contribution to total emission as a function of $J_{\mathrm{u}}$
within a 9${\arcsec}$ beam (1260 AU at 140 pc) for the three different models. 
The different colors correspond to different time steps: $t/t_{\mathrm{acc}}
\sim 0.50$ (red), 0.75 (blue), 0.96 (black).  }
 \label{fig:evoldiskcont}
\end{figure}

Figure~\ref{fig:evol1} shows the mass evolution of the evolutionary models
which indicates the times at which the models enter various evolutionary
stages.  Figure~\ref{fig:allc18o9} presents the \ceo\ 9--8 lines for the three
different evolutionary models convolved to a 9$\arcsec$ beam.  Most of the
line is emitted from the inner envelope region.  The line is generally broader
in later stages due to a combination of thermal, turbulent width and velocity
structure. However, the width of the line depends on whether the emitting region
is infall or rotation dominated.  Figures~\ref{fig:13colines}
and \ref{fig:evollines} shows the $J_{\rm u} =$3, 6 and 9 line profiles for the
different models and how they change within different beams. The disk
contribution to the observed lines within a 9$\arcsec$ beam is presented in
Fig.~\ref{fig:evoldiskcont} which is close to negligible for the low-$J$ lines.

%%%%%%%%%%%%%%%%%%%%%%%%%%%%%%%%%%%%%%%%%%%%%%%%%%%%%%%%%%%%%%%%%%%%%
%%%%%%%%%%%%%%%%%%%%%%%%%%%%%%%%%%%%%%%%%%%%%%%%%%%%%%%%%%%%%%%%%%%%%

\section{NIR molecular lines}\label{app:C}

\begin{figure}
 \centering
\raisebox{0.20cm}{\begin{minipage}[c]{0.95\linewidth}
\resizebox{\hsize}{!}{\includegraphics[angle=0,bb=35 180 560
580, clip]{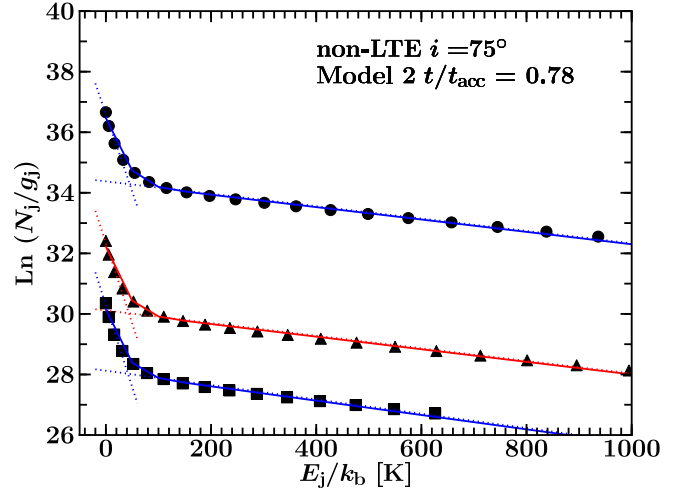}}
\end{minipage}} 
\caption{Example of the two-temperature fitting of the
  NIR lines.  The column densities are derived from a
  curve-of-growth analysis performed on the simulated spectra using
  RADLite.  The different symbols correspond to \mco\ (circle),
  \tco\ (triangles) and \ceo\ (squares).  The solid lines are the
  combination of the cold and warm components. }
 \label{fig:rotdiag1}
\end{figure}

\subsection{Inner disk}\label{app:C1}

\begin{figure}
 \centering
 \raisebox{0.20cm}{\begin{minipage}[c]{1.0\linewidth}
\resizebox{\hsize}{!}{\includegraphics[angle=0,bb=25 217 610 575, clip]{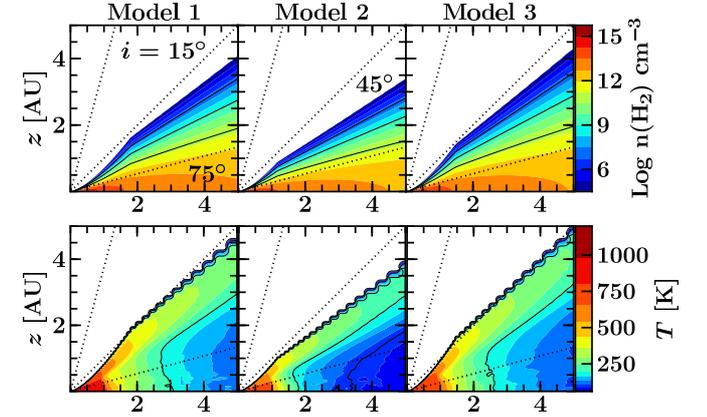}}
\end{minipage}} 
\caption{Density structure in the inner 10 AU (starting from n(H$_2$) = 
$10^{4.5}$ cm$^{-3}$) and temperature structure (starting from 60 K) of the
three models.  The three viewing angles are shown by dotted lines. }
 \label{fig:zoom}
\end{figure}

The formation of NIR continuum and the region where absorption happens depend on
the physical structure in the inner few AU as will discussed extensively in the
next section.  Figure~\ref{fig:zoom} shows the inner 5 AU structure of the three
different models.  In both Models 1 and 3, the $10^{4.5}$ cm$^{-3}$
density contour extends up to the line of sight of 45$^{\circ}$ due to the
presence of compact disk and massive disk, respectively.  On the other hand,
the density structure is flatter in Model 2.  Thus, the NIR continuum emission
is emitted through the high temperature region (mostly the red region in the
temperature plot) and encounters more material along the line of sight to an
observer in Models 1 and 3.  Consequently, the emission from the region close to
the midplane of the disk will not contribute to the NIR continuum.  
Figure~\ref{fig:h2col1} presents an example of the radial contribution to the
column density along a line of sight from the star.  At viewing angles
$<45^{\circ}$,  most of the warm material is located at 10--40 AU from the star
along the line of sight while most of the cold material is located $> 50$ AU.
The inner disk and high density region ($<1$ AU) is only accessible through
$i\ge75^{\circ}$ in our models. 

\subsection{Radiative transfer effects}\label{app:C2}

\begin{figure}
 \centering
\raisebox{0.20cm}{\begin{minipage}[c]{1.00\linewidth}
\resizebox{\hsize}{!}{\includegraphics[angle=0,bb=25 190 575
595, clip]{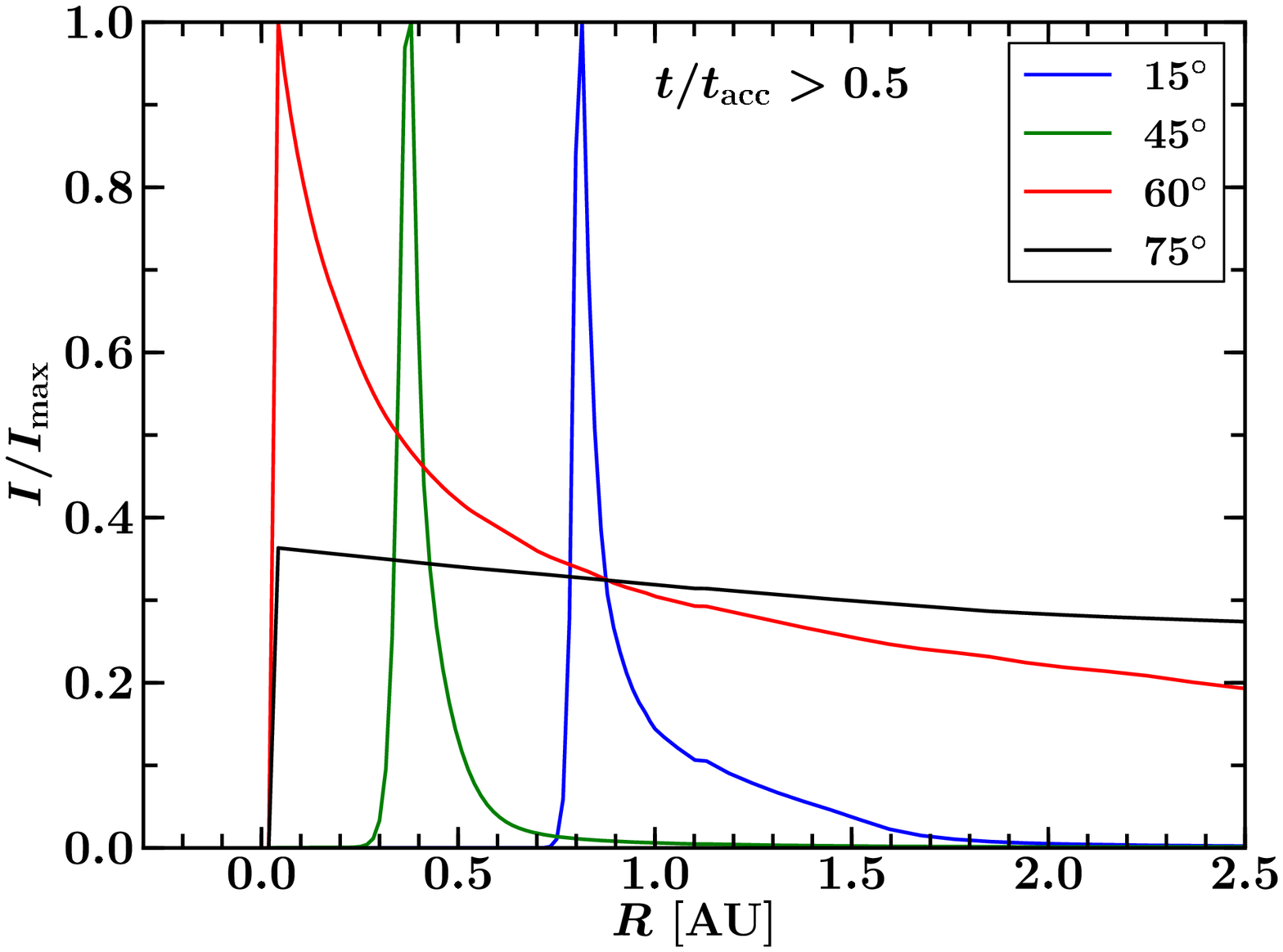}}
\end{minipage}}
\caption{Normalized continuum flux without the stellar photosphere as function
of distance from the center along one direction.  The figure is constructed
from a NIR image of Model 1 at $t/t_{\rm acc} = 0.7$ as an example.  The general
trend is similar for $t/t_{\rm acc} > 0.5$ independent of evolutionary models.
The normalized continuum is similar for the various evolutionary models for
$t/t_{\rm acc} > 0.5$. The continuum is shown to arise in the inner few AU
region and to fall off very quickly for $i < 60^{\circ}$. However, significant
contributions from larger radii are expected for relatively high inclination.}
\label{fig:contimage}
\end{figure}

Section~\ref{sec:nir} discusses the RADLite results with respect to the
predicted observables.  In this section, we present the comparison between
RADLite and a simple calculation using the integrated column density
(column method, equation~\ref{eq:tau0}) to examine the radiative transfer
effects in particular where the 4.7 $\mu$m continuum is formed and its affect
on the simulated absorption lines.

\begin{figure}
 \centering
\raisebox{0.20cm}{\begin{minipage}[c]{1.00\linewidth}
\resizebox{\hsize}{!}{\includegraphics[angle=0,bb=10 180 550 575, clip]{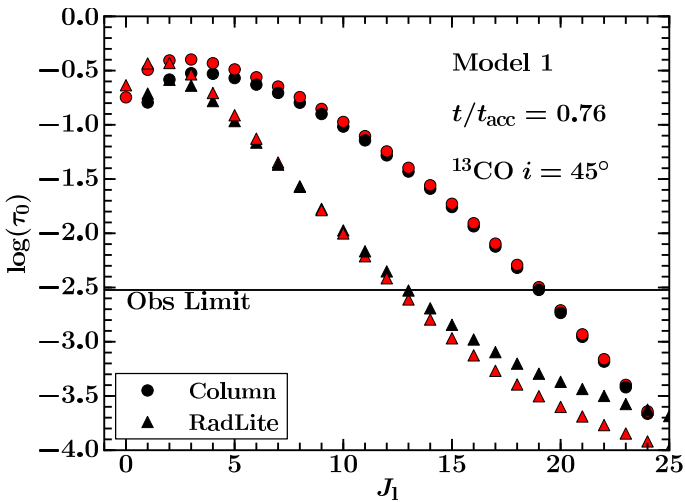}}
\end{minipage}} 
\caption{Comparison of the line center optical depth obtained from a simple
integration of the column density along the line of sight (circles) and the
radiative transfer calculation (triangles).  The $P$ and $R$ branches are shown
in black and red, respectively.  The horizontal line indicates the current
observation limit with CRIRES at $\tau \sim 3 \times 10^{-3}$.  The \tco\ lines
are shown for Model 1 at $t/t_{\rm acc} = 0.76$ for $i = 45^{\circ}$.}
 \label{fig:tau0comp}
\end{figure}

Figure~\ref{fig:tau0comp} illustrates the difference in line center optical
depth between the formal solution of the radiative transfer equation and the
column density approach at $i = 45^{\circ}$ (Eq.~\ref{eq:tau0}).  This
particular model and time was chosen as an example that clearly shows the
difference.  The difference in $\tau_0$ between the methods generally increases
with $J_{\rm l}$ and are largest for $J_{\rm l} \sim 10$.   Figure~\ref{fig:contimage} shows the 4.7 $\mu$m continuum flux due to dust only (i.e., no stellar contribution) as function of radius at various viewing angles.  The continuum is point-like for low inclinations but it is generally off-center for high inclinations.  Thus, larger $\tau_0$ differences between
the two methods are expected for larger inclinations since the absorption
does not take place directly along the line of sight to the star.  The
implication is that the column method overestimates $\tau_0$ of the
high-$J$ absorptions as this method takes the full line of sight through  the
warm material into account.  

What are the implications for translating observed optical depths to column
densities and excitation temperatures? The differences are largest for $J \ge 5$
lines which result in lower derived warm temperature and column
densities in the full treatment.  Consequently, the direct
transformation from the observed optical depth to column density requires
additional information on where the absorption arises, i.e., the NIR
continuum image. One way is to construct the physical structure and assess the
NIR continuum image.  Observationally, a comparison between interferometric
submm and NIR continuum should result in different continuum position if the
NIR continuum arises from scattered light.  Observationally driven
constraint is likely more useful in practice since modelling of the NIR
continuum requires a sophisticated inner disk and envelope physical structure.
For systems with an off-centered NIR continuum, the derived column densities
and temperature of the warm component are lower
limits.

\subsection{Non-LTE effects}\label{app:C3}

\begin{figure}
 \centering
\raisebox{0.20cm}{\begin{minipage}[c]{1.00\linewidth}
\resizebox{\hsize}{!}{\includegraphics[angle=0,bb=10 180 550 580, clip]{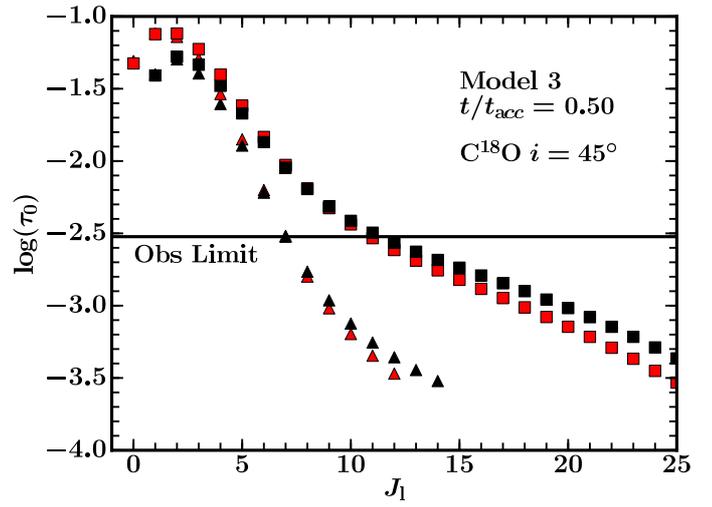}}
\end{minipage}} 
\caption{Comparison between LTE (squares) and non-LTE (triangles) line center
optical depth obtained from RADLite for Model 3 at an inclination of
45$^{\circ}$.  The different colors correspond to the $P$ (black) and $R$ (red)
branches. }
 \label{fig:nirtau0}
\end{figure}

The effects of non-LTE excitation in the vibrational ground state are studied by
either using the level populations calculated with the full non-LTE escape
probability method (\S 3) or assuming LTE populations. The LTE level population
is calculated using the partition functions provided by the HITRAN database
\citep{hitran}.

\begin{figure}
 \centering
\raisebox{0.20cm}{\begin{minipage}[c]{1.00\linewidth}
\resizebox{\hsize}{!}{\includegraphics[angle=0,bb=10 180 550 580, clip]{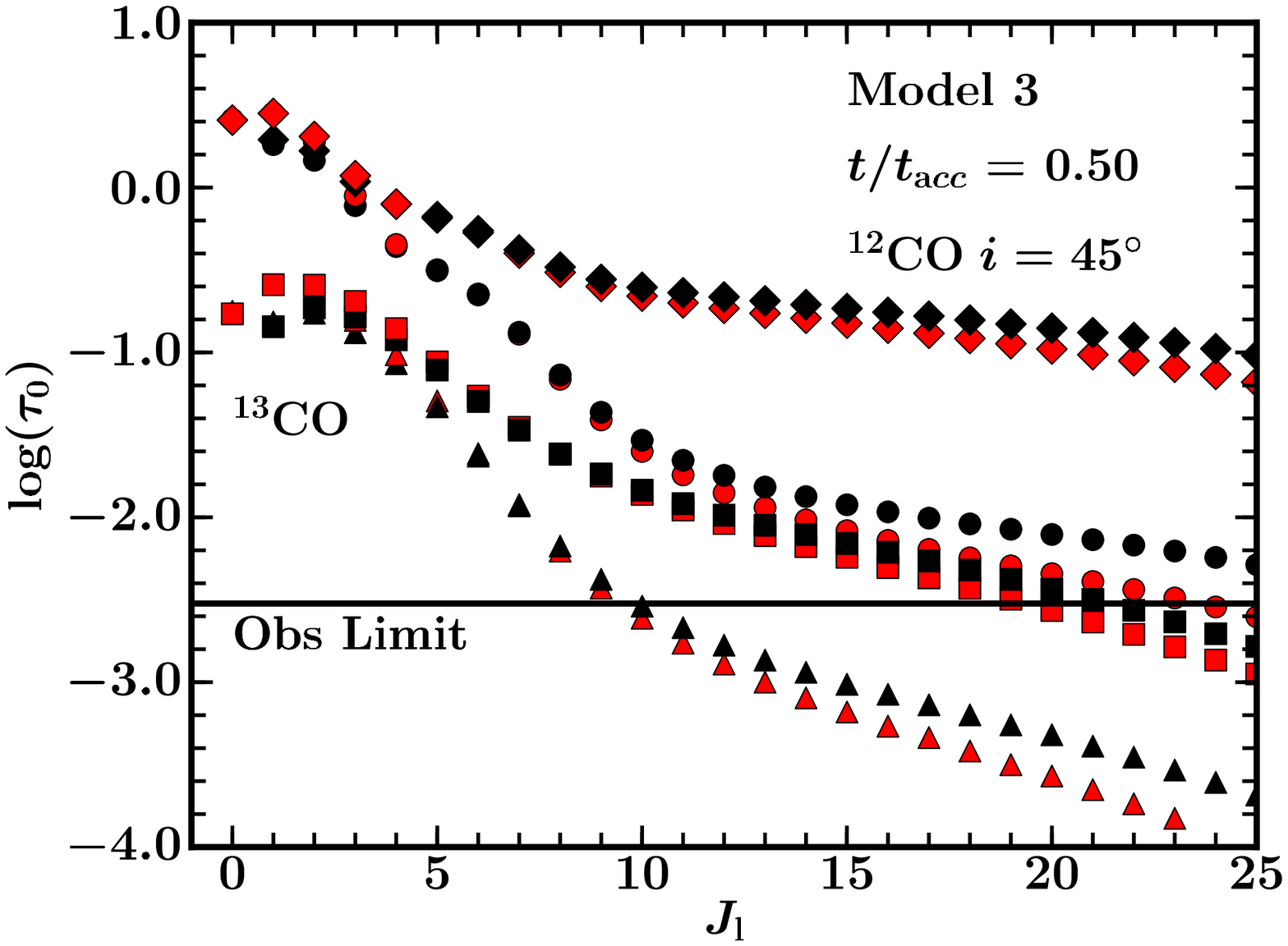}}
\end{minipage}} 
\caption{Similar to Fig.~\ref{fig:nirtau0} for \mco\ (diamonds and circles) and
  \tco\ (squares and triangles).}
 \label{fig:nirtau0a}
\end{figure}

Figure~\ref{fig:nirtau0} compares LTE and non-LTE line center opacities of
C$^{18}$O for Model 3 at an inclination of 45$^{\circ}$ (other isotopologs
are shown in Figure~\ref{fig:nirtau0a}).  Model 3 was chosen as an example; in
general, the non-LTE effects are most apparent for the higher levels independent
of isotopologs and evolutionary models.  This reflects the findings of \S 3:
lower pure rotational levels are more easy to thermalize due to lower critical
densities.  The difference between LTE and non-LTE is independent of
inclination while radiative transfer effects do.  Thus, it is more important to
treat the radiative transfer correctly in the NIR to derive observables.

%%%%%%%%%%%%%%%%%%%%%%%%%%%%%%%%%%%%%%%%%%%%%%%%%%%%%%%%%%%%%%%%%%%%%%%%%%%%%%%%%%%%%%%%%%%%%%%%%%%%%%%%%%%%%%%%%%%%%
%%%%%%%%%%%%%%%%%%%%%%%%%%%%%%%%%%%%%%%%%%%%%%%%%%%%%%%%%%%%%%%%%%%%%%%%%%%%%%%%%%%%%%%%%%%%%%%%%%%%%%%%%%%%%%%%%%%%%

\end{document}